\documentclass[traditabstract]{aa} 
\usepackage[varg]{txfonts}
\usepackage{color}
\usepackage{multirow}
\usepackage{graphicx}
\usepackage{gensymb}
\usepackage{epstopdf}
\usepackage{appendix}
\usepackage{siunitx}
\usepackage{textcomp}
\usepackage{amsmath}
\usepackage{comment}

\usepackage{natbib}
\bibpunct{(}{)}{;}{a}{}{,} 

\newcommand{\Msub}[1]{\ensuremath{M_{\mathrm{#1}}}}

\newcommand{\hst}{\emph{HST}}

\begin{document}

\title{Tracing the anemic stellar halo of M101}


\author{In Sung Jang\inst{1}
  \and Roelof S. de Jong\inst{1} 
     \and Benne W. Holwerda\inst{2}
     \and Antonela Monachesi\inst{3,4}
     \and Eric F. Bell\inst{5}
     \and Jeremy Bailin\inst{6}}


\institute{Leibniz-Institut f\"{u}r Astrophysik Potsdam (AIP), An der Sternwarte 16, 14482 Potsdam, Germany
  \and Department of Physics and Astronomy, 102 Natural Science Building, University of Louisville, Louisville KY 40292, USA
  \and Instituto de Investigaci\'on Multidisciplinar en Ciencia y Tecnolog\'ia, Universidad de La Serena, Ra\'ul Bitr\'an 1305, La Serena, Chile
  \and Departamento de F\'isica y Astronom\'ia, Universidad de La Serena, Av. Juan Cisternas 1200 Norte, La Serena, Chile
   \and Department of Astronomy, University of Michigan, 311 West Hall, 1085 South University Ave., Ann Arbor, MI 48109-1107, USA
   \and Department of Physics and Astronomy, University of Alabama, Box 870324, Tuscaloosa, AL 35487-0324, USA}


\abstract{Models of galaxy formation in a cosmological context predict that massive disk galaxies should have structured extended stellar halos.
 Recent studies in integrated light, however, report a few galaxies, including the nearby disk galaxy M101, that have no measurable stellar halos to the detection limit. 
We aim to quantify the stellar content and structure of M101's outskirts by resolving its stars.
We present the photometry of its stars based on deep F606W and F814W images taken with \emph{Hubble Space Telescope} (\hst) as part of the GHOSTS survey.
The \hst\  fields are placed along the east and west sides of M101 out to galactocentric distance ($R$) of $\sim$70 kpc.
The constructed color-magnitude diagrams of stars reach down to two magnitudes below the tip of the red giant branch.
We derived radial number density profiles of the bright red giant branch (RGB) stars.
The mean color of the RGB stars at $R \sim$ 40 -- 60 kpc is similar to those of metal-poor globular clusters in the Milky Way. 
We also derived radial surface brightness profiles using the public image data provided by the Dragonfly team. 
Both the radial number density and surface brightness profiles were converted to radial mass density profiles and combined.
We find that the mass density profiles show a weak upturn at the very outer region, where surface brightness is as faint as $\mu_g\approx 34$ mag arcsec$^{-1}$.
An exponential disk + power-law halo model on the mass density profiles finds the total stellar halo mass of $M_{halo}=8.2_{-2.2}^{+3.5}\times 10^7M_\odot$.
The total stellar halo mass does not exceed $\Msub{halo} = 3.2 \times 10^8$ $M_{\odot}$ 
when strongly truncated disk models are considered.
Combining the halo mass with the total stellar mass of M101,
  we obtain the stellar halo mass fraction of $\Msub{halo}/\Msub{gal} = 0.20_{-0.08}^{+0.10}\%$ with an upper limit of 0.78\%.  
We compare the halo properties of M101 with those of six GHOSTS survey galaxies as well as the Milky Way and M31 and find that M101 has an anemic stellar halo.}

\maketitle

\section{INTRODUCTION}

The standard paradigm of the $\Lambda$CDM cosmology predicts the hierarchical assembly of stellar and dark matter halos during structure formation \citep{whi78,blu84}. 
Massive galaxies, either elliptical or disk galaxies, have grown through the accretion of gas and the merger of preexisting less massive galaxies.
The latter process leads to the formation of extended stellar envelopes around massive galaxies, called stellar halos.
Because the dynamical and star formation time-scales at galaxy outskirts are considered to be long, stellar halos are ideal systems in which to study the growth and assembly history of galaxies \citep{bul01, fre02, bul05}.

The GHOSTS\footnote{GHOSTS: the Galaxy Halos, Outer disks, Substructure, Thick disks and Star clusters} survey is an ongoing \emph{Hubble Space  Telescope} (\hst)
survey program, which images the outskirts of nearby galaxies to a provide better understanding of the disk growth and the halo assembly histories through resolved stars \citep{dej07, rad11}.
Our recent studies based on six Milky Way (MW)-mass disk galaxies have shown that these six galaxies as well as the two massive galaxies in the Local Group (the MW and M31) have stellar halos dominated by old and metal-poor red giant branch (RGB) stars and they show large variations in mass, mass fraction, median color (metallicity), color gradient, and power-law surface brightness profile slope \citep{mon16, har17}.
All these properties agree well with the predictions of cosmological simulations that firstly, massive galaxies should have stellar halos, and secondly, their properties should vary considerably from galaxy to galaxy. 
\citep{bul01, bul05, pur07, coo10, mon19}.

It is therefore puzzling that recent studies in integrated light with a novel array of
telephoto lenses 
(the Dragonfly array \citep{abr14}) 
reported a few galaxies that have no measurable stellar halos to faint limits ($\mu_g \sim 32$ mag arcsec$^{-2}$).
Van Dokkum et al. (2014) investigated the outskirts of the massive spiral galaxy M101 with the Dragonfly array. 
By fitting the mass density profile with a multicomponent model, they find the stellar halo mass fraction, $\Msub{halo}/\Msub{gal} = 0.3^{+0.6}_{-0.3} \%$, to be significantly lower than those of other massive disk galaxies ($\Msub{halo}/\Msub{gal}$ $\simeq$ 1 to a few $\%$).
The stellar halo of M101 was revisited by the Dragonfly team in \citet{mer16}. 
From the same integrated light image data, they measured the stellar halo mass outside of five half-mass radius ($R_h$)
and derived a value of the halo mass fraction, $M_{\mathrm{halo}, >5R_h}/\Msub{gal} = 0.04\pm0.08 \%$.
Although they did not explicitly derive the total stellar halo mass fraction, it is expected to be several times larger than the $>5R_h$ value \citep{har17}, which would agree with \citet{van14}.
Considering the low halo mass fractions and large uncertainties, the very presence of an M101 stellar halo is controversial.
\citet{mer16} also report two more galaxies that have no detectable stellar halos: 
NGC 1042 ($\Msub{halo,>5R_h}/\Msub{gal} = 0.01\pm0.01 \%$) and 
NGC 3351 ($\Msub{halo,>5R_h}/\Msub{gal} = 0.02\pm2.25 \%$).
The low inclinations (close to face-on) of these three galaxies makes detecting a stellar halo very challenging, but taken at face value, such a population of galaxies without stellar halos would present a critical challenge to the picture of ubiquitous stellar halos formed from the debris of disrupting dwarf galaxies.

Stellar halos are in general very faint so a quantitative measurement of the halo component in a galaxy requires intensive efforts.
Models predict that most of the stars in a stellar halo are at small galactocentric radii, where they are very difficult to disentangle from the much brighter bulge and disk components. Stellar halos are dominant only at larger radii, where the surface brightnesses can reach well below $\mu \sim 30$ mag arcsec$^{-2}$.
Integrated light photometry is a useful approach for detecting stellar halos owing to surface brightness being distance independent and the relatively modest observational cost for covering large areas of sky. 
Yet, detecting and quantifying halo light up to 10000$\times$ fainter than the sky requires stringent control of systematic error. 
A small change in the background modeling, foreground star or background galaxy subtraction can lead to a large difference in the derived halo profile.
The scattered light from the optics \citep{dej08, sla09, abr14, san14} or the atmospheric conditions \citep{dev13, zha18} and low surface brightness galactic cirrus \citep{tru16} are  other possible contaminants, which make the measurement of the low surface brightness stellar structures using integrated light more difficult and much more uncertain.

A complementary and more reliable approach is to count and characterize resolved stars in order to quantify the halo structure.
It enables to obtain information of age- and metallicity-resolved stellar populations, which allows clearer separation between the stellar disk and halo, and exploration of variations in stellar population from halo to halo \citep{mon16, har17}.
While resolved stars from wide field ground based imaging is possible for the nearest galaxies (e.g., M31 - \citet{iba14}, NGC 253 - \citet{gre14}, Cen A - \citet{crn16}, M81 - \citet{oka15}), star counting with $HST$ is able to 
detect a faint stellar structure that has surface brightness levels down to $\mu_V \approx 32 - 34$ mag arcsec$^{-2}$ \citep{str16, har17}, a factor of roughly ten deeper than limits achieved with integrated light, $\mu_V \approx 30 - 32$ mag arcsec$^{-2}$ \citep{mih13, mih17, mer16}. 
These limits can even be achieved on relatively small patches on the sky, without having to average over very large annuli to obtain enough signal to noise, and is mainly limited by the Poisson noise in the number of detected objects and the statistical removal of contaminating sources. The primary disadvantages of resolved star studies are that they are strongly limited in distance to galaxies within $\sim 10-15$\,Mpc \citep{rad11, lee16} and that coverage with \hst\  or many other facilities is limited to small fields of view, making it difficult to survey large areas. 
These two independent approaches each have their own pros and cons in detecting faint stellar substructures, and it would be ideal to combine both techniques whenever practical.

In this study, we select M101, a face-on massive disk galaxy, to study its faint outskirts based on both star count and integrated light photometry. 
M101 has similar properties to the MW in terms of absolute magnitude ($M_V \sim -21.0$ mag \citep{mak14}), maximum rotation velocity ($V_{max} \sim 235$  $km s^{-1}$ \citep{bos81, zar90}), and morphological type (SABcd \citep{dev91}).
There are $\sim$10 known dwarf galaxies within 100 kpc projected distance from M101 \citep{kar15, dan17, ben19}, possibly associated with the formation of the M101 outskirts. 
M101 has globular clusters, which show the global spatial distribution similar to the MW globular clusters \citep{sim15}. 
The line-of-sight velocities of the M101 globular clusters do not follow the HI disk kinematics, so some of them could be located well above the disk plane,  being counterparts of the halo globular clusters in the MW \citep{sim17}.

The lack of a detectable stellar halo in M101 is hard to understand given satellite and globular cluster populations.
While numerical simulations predict that halos of MW mass galaxies can have a wide range of masses, halos are inevitable when galaxies are formed through mergers of smaller systems \citep{pur07, pur08, dea16, dso18}.
If the lack of stellar in M101 is taken at face value it would be 
impossible to interpret, and would indicate and important flaw in hierarchical models. 
Consequently, we study M101's distant outskirts using resolved stars in concert with public integrated light image data, to provide a much more robust measurement of its outskirts properties.

The distance modulus to M101 is estimated to be $(m-M)_0 = 29.11 \pm 0.04$ ($d = 6.64\pm0.12$ Mpc) from recent studies based on near-infrared Cepheids \citep{rie16} and optical Tip of the Red Giant Branch (TRGB) measurements \citep{bea19, fre19}. At this distance, its disk and halo can readily be resolved into individual stars through \hst\  imaging with reasonable exposure times.

The stellar halo of M101 is, however, far more difficult to be detected compared to other massive galaxies because of its face-on nature \citep[$incl. \approx 17\degree$, i.e.,][]{zar90}. 
The stellar disk of M101 extends in radius well beyond $15\arcmin$ (30 kpc, see Figure \ref{fig_finder1}) with a mean disk scale length of $R_d \approx 2\farcm2$ (4.2 kpc) \citep{mih13, lai16}, which is almost a factor of two larger than that of the MW, $R_d \approx 2.2$ kpc \citep{bov13}.
M101 is a textbook example of a lopsided galaxy \citep{jog09}, presenting a strong asymmetry of the outer disk structure.
Its disk scale length varies significantly with azimuthal angles from $1\farcm9$ (3.7 kpc) to $3\farcm4$ (6.6 kpc) \citep{mih13}. 
The stellar halo of M101, if any, is expected to be embedded in the extended, asymmetric, and face-on stellar disk and to dominate the luminosity only at the very outer regions. 
The resolved stellar populations can help to detect stellar halo in this galaxy, 
providing a much more robust measurement of its stellar halo properties.

This paper is organized as follows.
In Section 2 we describe the data we used and how we derive the stellar photometry.
Section 3 presents color-magnitude diagrams (CMDs) of target fields. 
We show radial star count profiles for young, intermediate, and old stellar populations in M101.
We also present integrated light photometry based on the public images provided by the Dragonfly Nearby Galaxy Survey.
Then we combine both the star count and integrated light profiles and derive the mass contribution of the stellar halo.
Section 4 discusses the halo properties of M101 in comparison with those in previous studies, the other observed halos, and model predictions.
Primary results are summarized in Section 5.

\section{OBSERVATIONS AND DATA REDUCTION}

\subsection{Observations}
\begin{figure*}
\centering
\includegraphics[scale=1.6]{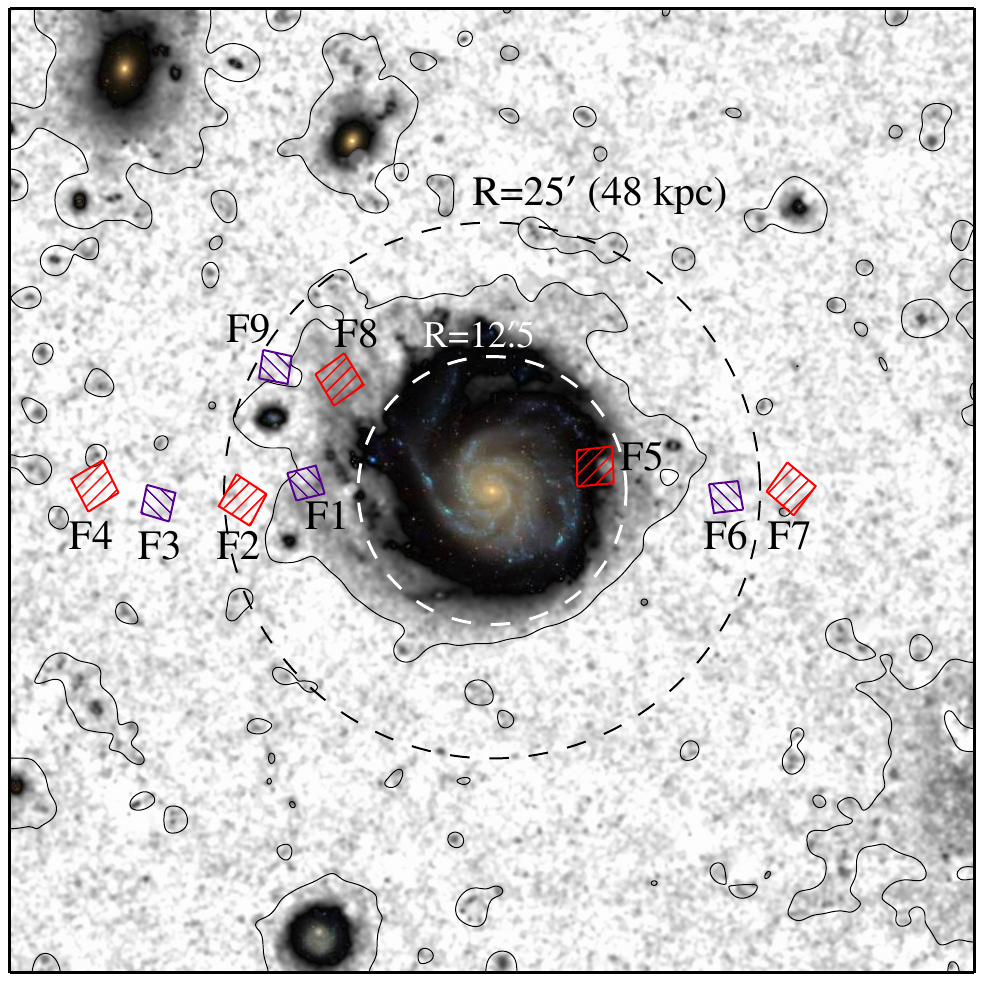}
\caption{Location of \hst\  ACS (red squares) and WFC3 (purple squares) fields overlaid on a deep $75\arcmin \times 75\arcmin$ g-band image provided by the Dragonfly project \citep{abr14}. 
North is up and east is to the left.
High surface brightness regions of the $g$-band image are replaced by the color image from the SDSS.
The white and black dashed lines represent circles with $12\farcm5$ (24 kpc) and $25\arcmin$ (48 kpc) radii centered on M101, respectively.
A black contour line indicates a smoothed iso-surface brightness level of $\mu_g = 30$ mag/arcsec$^2$ measured in this study.
}
\label{fig_finder1}
\end{figure*}

The observations were taken between 14 February and 9 September 2015 with \hst\  under the program ID 13696 (PI: B. Holwerda) 
as part of the GHOSTS survey (PI: R. de Jong). 
Six target fields were chosen to sample resolved stellar populations at the outskirts of M101 along the east and west sides from the center, as shown in Figure \ref{fig_finder1} (F1 to F7 except for F5). 
The six fields cover $R = 17\farcm4$ to $37\farcm0$, corresponding to galactocentric distances $R$ = 34 to 71 kpc from the M101 center. 
Two broadband filters, $F606W$ and $F814W$ in the Advanced Camera for Surveys (ACS) and the Wide Field Camera 3 (WFC3), have been used for simultaneously measuring the density via star count and metallicity via color of the RGB following \citet{mon13, mon16} and \citet{har17}.
Exposure times are $\sim$2700s for each filter, long enough to detect resolved RGB stars with a signal to noise ratio of 10 at $\sim$ 2 mag below the TRGB. 
We also used three \hst\  fields in the archive: 
one shallow ACS field (F5) located at $R$ = $9\farcm9$  from the galaxy center (ID=13364, PI: D. Calzetti) and two very deep fields (F8 and F9) taken with ACS and WFC3 
located at the northeast side of M101 (ID=13701, PI: C. Mihos).
A summary of the \hst\  data used in this work for M101 is listed in Table \ref{tab_obs1}.\\

\begin{table*}
\caption{Summary of HST data for M101 (NGC 5457)}
\centering
\begin{tabular}{lcccrrrr}
\hline
\hline
Field & R.A. & Decl & $R$ & Inst. & \multicolumn{2}{c}{Exposure Time(s)} & Prop ID  \\
 & (2000.0) & (2000.0) & [arcmin (kpc)$^a$] & & $F606W$  & $F814W$  \\
\hline
F1	& 14 05 11.64 & 54 21 35.5	& 17.4 (34)	& WFC3	& 2,776 & 2,896 & 13696 \\ 
F2	& 14 05 51.65 & 54 19 59.3	& 23.2	(45)	& ACS	& 2,540 & 2,678 & 13696 \\ 
F3  & 14 06 46.21 & 54 19 36.0	& 31.2	(60) 	& WFC3	& 2,776 & 2,896 & 13696 \\ 
F4	& 14 07 26.34 & 54 21 07.5	& 37.0	(71)	& ACS	& 2,540 & 2,678 & 13696 \\ 
F5	& 14 02 06.53 &	54 23 11.6	& 9.9 (19) 	& ACS	& 1,130 & 1,420 & 13364 \\   
F6	& 14 00 42.59 & 54 20 17.6	& 21.9 (42)	& WFC3	& 2,776 & 2,896 & 13696 \\ 
F7	& 14 00 01.40 &	54 20 59.5	& 27.9 (54)	& ACS	& 2,540 & 2,678 & 13696 \\ 

F8	& 14 04 50.66 &	54 31 16.6	& 17.7 (34) 	& ACS	&38,290 & 32,820 & 13701 \\   

F9	& 14 05 31.62 & 54 32 22.0	& 23.3 (45) 	& WFC3	&34,200 & 39,900 & 13701 \\   
\hline
\end{tabular}
\tablefoot{\\
\tablefoottext{a}{A face-on (symmetric) geometry assumed.}
}
\label{tab_obs1}
\end{table*}

\subsection{Data reduction and photometry}
The data reduction and the photometry were performed using the GHOSTS pipeline. 
The pipelines for the ACS/WFC and WFC3/UVIS data are described in detail in \citet{rad11} and \citet{mon16}, respectively.
We briefly summarize the general procedure and minor changes we applied to the pipeline in this study.

The pipeline is composed of three main steps: source detection, stellar photometry, and point source selection.
The first and second steps are based on DOLPHOT \citep{dol00}, and the last step utilizes SExtractor \citep{ber96}, as well as photometric culls.
All these steps are executed automatically by Python-based routines. 

The source detection was carried out on the combined, drizzled images.
We downloaded the charge transfer efficiency-corrected individual frame images (FLC images) of ACS/WFC and WFC3/UVIS observations from MAST$\footnote{https://mast.stsci.edu}$. 
We noted that some of the FLC images are not properly aligned, showing small offsets of $\sim0.5$ pixel, so that the drizzled images have slightly broader PSFs.
We made a bright source list for each FLC image and updated the WCS in the image header with Tweakreg. 
The resulting astrometry was good to 0.1 pixel.
We then ran astrodrizzle to make drizzled images. 
In this process, data quality (DQ) extension images were also updated.

\begin{figure*}
\centering
\includegraphics[scale=0.9]{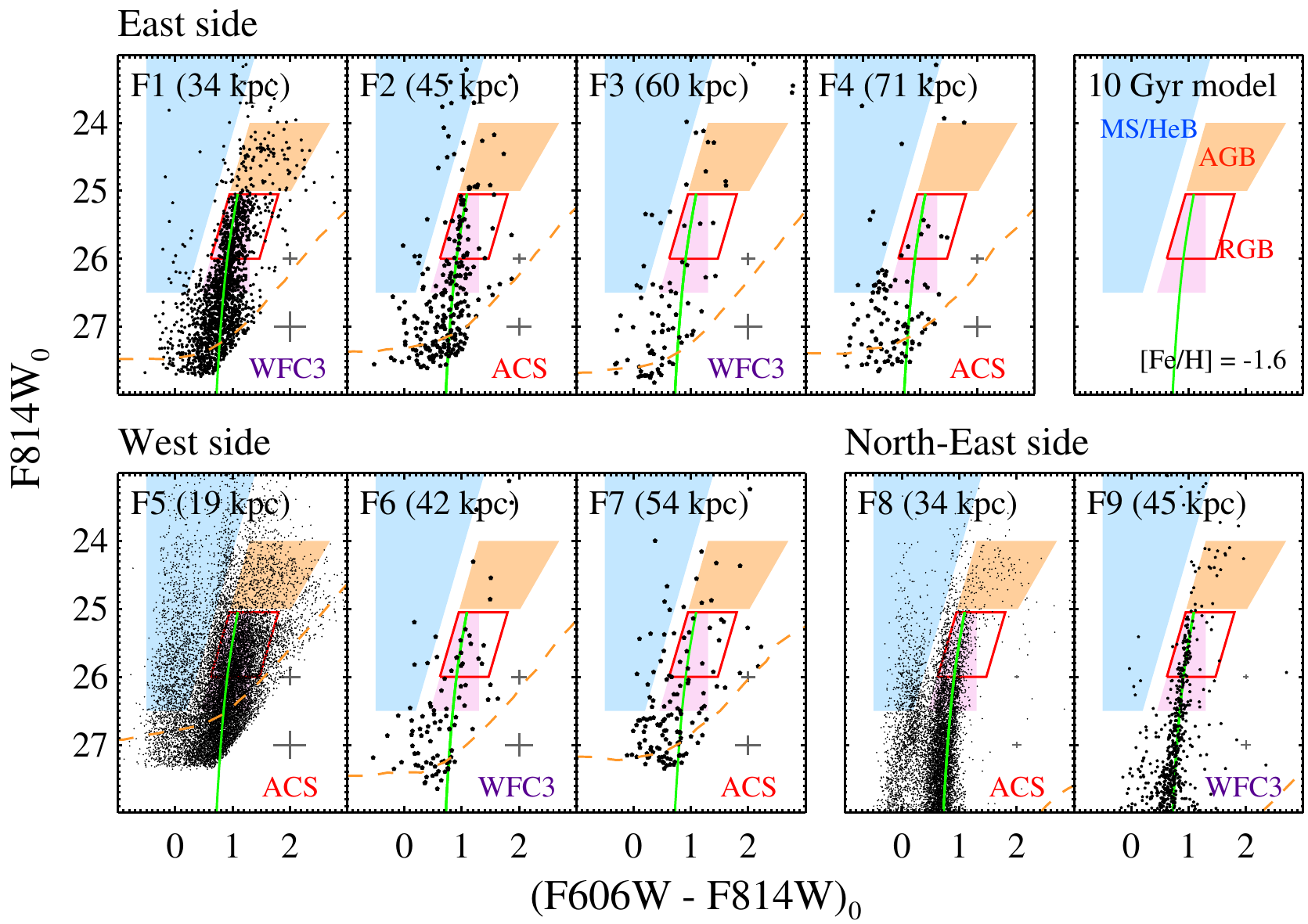}
\caption{ 
CMDs for the selected point sources corrected for Galactic extinction and reddening in the nine $HST$ fields around M101. 
Blue, orange, and pink shaded regions indicate selection bins for the young (main sequence and helium burning stars), intermediate (asymptotic giant branch stars), and old (RGB stars) stellar populations, respectively. 
The region enclosed by red lines represent the selection bin we used for the color measurement only.
Dashed lines represent the 50\% completeness limits. 
Photometric errors as derived from ASTs at $(F606W - F814W)_0$ = 1.0 are marked at $(F606W - F814W)_0$ = 2.0.
We overlaid a 10 Gyr stellar isochrone with [Fe/H] = --1.6  in the Padova models \citep{bre12}, to give an indication of the RGB domain in the CMD. 
\label{fig_cmd1} 
}
\end{figure*}

The stellar photometry was carried out using the ACS and WFC3 modules of DOLPHOT on the WCS-corrected FLC images. 
Cosmic rays and hot pixels in the science extension of the FLC images were masked using the DQ extension.
We used TinyTim PSFs \citep{kir11} implemented in DOLPHOT for the ACS reduction. In the case of the WFC3 reduction, we use Jay Anderson's PSFs \citep{and06}, because we found that they provide better photometric measurements showing smaller systematic offsets in overlapping regions of WFC3 images \citep{mon16}.
The DOLPHOT parameters used for the ACS/WFC and WFC/UVIS reductions in this study are the same as those given in Table A2 of \citet{mon16}. 

The raw DOLPHOT output catalogs contain various types of sources, such as background galaxies, saturated stars, artifacts, as well as point sources. 
We selected the clean point sources by applying several selection criteria.
We first applied photometric culls as follows:

\begin{equation}
-0.06 < \mathrm{SHARPNESS_{F606W}} + \mathrm{SHARPNESS_{F814W}} < 1.30
\end{equation}
\begin{equation}
\mathrm{CROWDING_{F606W}} + \mathrm{CROWDING_{F814W}} < 0.16
\end{equation}
\begin{equation}
\mathrm{S/N_{F606W} > 5.0}, \qquad \mathrm{S/N_{F814W} > 5.0}
\end{equation}
for the ACS/WFC reduction, and

\begin{equation}
-0.19 < \mathrm{SHARPNESS_{F606W}} + \mathrm{SHARPNESS_{F814W}} < 1.50
\end{equation}
\begin{equation}
\mathrm{CROWDING_{F606W}} + \mathrm{CROWDING_{F814W}} < 0.20
\end{equation}
\begin{equation}
\mathrm{S/N_{F606W} > 5.1}, \qquad \mathrm{S/N_{F814W} > 3.2}
\end{equation}
for the WFC3/UVIS reduction.
Additionally, we selected sources with type = 1 (good star). 
These culls were determined from the analysis of empty archival fields, which are dominated by background galaxies. 
Our previous studies found that these culls are very efficient in the point source selection, eliminating 90 -- 95\% of the background galaxies \citep{rad11, mon16}.
Even after the culls, there are some remaining background galaxies (5 -- 10\%). 
Characterizing their contribution to the point source photometry is also important to derive the true radial number density profile of the M101 stars.
We used extra empty fields to determine the likely background of the M101 data more accurately.
Details of the selected fields and the basic procedure of the background estimation are presented in Appendix~\ref{sec:appendix-controlfields}.

Second, we made a mask file of extended sources using SExtractor 
and rejected sources lying in the pixel position of the masked sources in the final catalog.
Finally, we visually inspected all the remaining sources brighter than F814W = 26.5 mag in five fields, F2, F3, F4, F6, and F7, which are expected to show very low stellar density, and confirmed that all the detected sources are genuine point-like sources. 
In the case of F7, however, we discovered a clustering of point sources centered at $\alpha$ = $13^h59^m46.69^s$ and $\delta$ = $+54\degree21\arcmin22\farcs32$, possibly a 
dwarf galaxy associated with M101 or a background galaxy
(Jang et al. in preparation). We discarded point sources in $R=3\arcsec$ region around the galaxy candidate in the following analysis.

The \hst\  data used in this study were taken with two instruments, ACS/WFC and WFC3/UVIS. 
We converted the WFC3/UVIS magnitudes to the ACS/WFC magnitude system using the photometric transformation in \citet{jan15}: 

\begin{equation}
F606W_{ACS} = F606W_{WFC3} + (0.002\pm0.002) - (0.032\pm0.002) C
\end{equation}
and
\begin{equation}
F814W_{ACS} = F814W_{WFC3} + (0.016\pm0.002) - (0.006\pm0.002) C
\end{equation}
where $C \equiv F606W_{ACS} - F814W_{ACS}$.
This empirical transformation is based on stars in two Milky Way globular clusters (NGC 2419 and 47 Tuc), which were observed with both ACS/WFC and WFC3/UVIS.
The photometry of resolved stars in this paper is given in the Vega magnitude system.

\subsection{Artificial star tests (ASTs)}

We carried out extensive ASTs to estimate the level of photometric completeness and uncertainties.
The procedure of the tests is described in detail in \citet{rad11}. 
In short, we placed a number of artificial stars in each frame and passed them through DOLPHOT. This process was repeated until we obtained in total $\sim$1,000,000 artificial stars per field.
We set the spatial distribution of the input artificial stars based on surface brightness, placing more stars in the higher surface brightness regions.
The distribution of color and magnitude of the artificial stars were similar to those of the real stars.

We found that the 50\% completeness levels for the RGB candidates in fields F1 to F7 except for F5 are approximately at $F814W$ = 27 mag, two magnitudes below the TRGB.
The same completeness level for Field 5 was reached at $F814W \simeq$ 26.4 mag, slightly brighter than those of the other fields, but still deep enough to investigate the RGB population. 
The low completeness of Field 5 is due to its high stellar density and shorter exposure time.
Fields F8 and F9 were taken with much longer exposure times than the other fields.
We obtained the deepest stellar photometry from these two fields such that the 50\% completeness levels reach down to $F814W$ = 28.3 mag, more than three magnitudes below the TRGB.

We also confirmed that our photometry is precise and accurate. 
The $1\sigma$ dispersion in magnitude is as small as 0.05 mag at $F814W$ = 25 mag 
in fields F1 to F7. The dispersion becomes slightly larger at $F814W$ = 26 mag, but still an order of 0.1 mag: $0.08 \sim 0.1$ mag for F1 to F7 except for F5, and $0.13$ mag for F5.
The two deep fields, F8 and F9, are much more precise than the other fields,
showing the dispersion smaller than 0.05 mag for all magnitude bins brighter than $F814W = 26.5$ mag.
Systematic offsets in photometry, which are the mean magnitude differences between the input sources and the recovered sources, are also measured to be small.
The offsets are smaller than 0.03 mag for stars brighter than F814W = 25 mag in all the fields.
The offsets do not exceed 0.07 mag at F814W = 26.5, well within the typical errors at the same magnitude level.

\begin{figure*}
\centering
\includegraphics[scale=0.9]{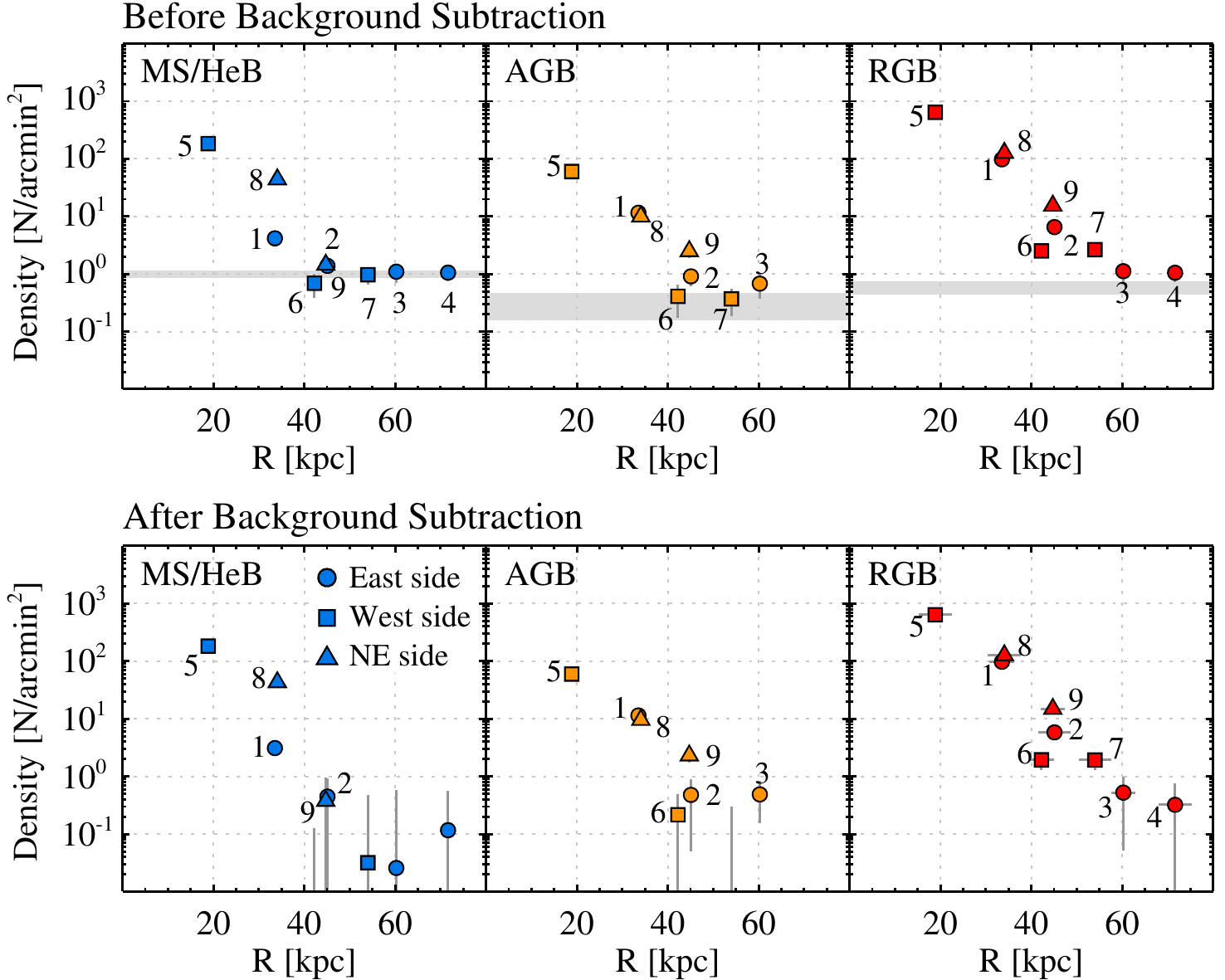}
\caption{(Top panels) Radial projected stellar density profiles 
for the MS/HeB (left), AGB (middle), and RGB (right) populations before background subtraction.
Circles, squares, and triangles with field IDs in each panel indicate the profiles along the east, west, and northeast sides from the center of M101, respectively.
Error bars indicate 1$\sigma$ Poisson uncertainties. 
The approximate background levels measured from the control fields are marked by shaded regions.
(Bottom panels) Same as the top panels, except after background subtraction.
In the last panel, we indicate the radial coverage of each field by a small horizontal line.
}
\label{fig_rdp4}
\end{figure*}

\section{RESULTS}

\subsection{Color magnitude diagrams (CMDs) of the M101 fields}
In Figure \ref{fig_cmd1} we display CMDs for the selected point sources in the nine fields of M101. 
The mean Galactocentric distances, assuming a face-on symmetric geometry, are marked in each panel.
The galactic foreground extinction toward M101 is known to be very low, $A_{F606W} = 0.021$ mag and $A_{F814W} = 0.013$ \citep[][NED]{sch11}, and the extinction of the nine fields are nearly the same.
The host galaxy extinction is expected to be very low, because most of the target fields cover the outermost regions of the galaxy. 
We corrected the magnitudes for the foreground extinction only.

CMDs allow us to investigate the age-resolved stellar population. 
We defined three CMD bins as indicated by shaded regions in each panel to preferentially sample young, intermediate, and old stellar populations.
The bluest CMD bin samples bright main-sequence and helium-burning stars, typically younger than 300 Myr (blue region).
The two CMD bins in the red color range are designed for asymptotic giant branch (AGB) stars with ages of between $\sim1$ and $\sim3$ Gyr (orange region), and old RGB stars, mostly older than 3 Gyr (pink region).
We set a narrow color range for the RGB bin, as fields F2 and F9 show already very narrow, blue (metal-poor) RGB distributions and fields at larger radii can be expected similarly metal-poor.
All the CMD bins, except for the bin sampling RGB stars in F5, are well above the 50\% completeness level (orange dashed line).
The CMD of F5 is somewhat shallower than the other fields due to the short exposure time and high stellar density in the inner region of M101.
Nevertheless, the CMD is deep enough to detect stars down to $\sim1.5$ mag below the TRGB and the applied incompleteness corrections are small.

The CMDs show a few notable features.
First, the CMDs for the inner region of M101 show signs of younger stellar populations, as well as old RGB stars. 
This is evident in F5, covering the innermost region at $R\sim 19$ kpc.
There is a plume of blue MS stars, and red helium burning (RHeB) stars, so F5 is estimated to be dominated by or significantly contaminated by disk stars. 
The CMD of F8 at $R\sim 34$ kpc shows similar features, 
but the mean RGB color is much bluer and the MS and RHeB stars in F8 do not reach as bright magnitude as in F5, thus indicating that F8 lacks the youngest populations seen in F5.
The CMD of F1  shows an evident population of AGB stars.
We noted that two fields, F1 and F8 cover extended faint structures of M101, called ``E-spur'' and ``NE plume'' in \citet{mih13}, which show bluer $B-V$ color with a stronger far-ultraviolet emission than other regions at the same galactocentric distance.
Thus, the sign of younger stellar populations in the two fields is consistent with the result of the integrated light study.

Second, CMDs for F2 and F9 show a well-defined narrow and blue RGB stellar population.
We find that the RGB sequence of this field is well matched to the 12 Gyr isochrone with a metallicity of [Fe/H] $\approx$ --1.6 in the Padova models \citep{bre12}.
F5 and F1 in the inner region at $R$ = 19 kpc and 34 kpc, respectively, show mostly redder and broader RGB stars compared to the other fields. 
In this sense, RGB stars appear to become bluer from inner to outer region (F5-F1-F8-F2-F9) as can be seen from their location relative to the isochrone shown in Figure \ref{fig_cmd1}.
Fields beyond 50 kpc (F3, F4, and F7) are expected to be dominated by metal-poor RGB stars, but it is not obvious in the CMDs due to their low stellar densities with underlying contaminants, mostly either foreground stars in the Milky Way or unresolved background galaxies.

Third, fields F2, F6, and F9 cover nearly the same projected distances along the east, west, and northeast sides of M101, respectively, but their RGB sequences in CMDs are distinctly different.
The CMDs of F2 and F9 have a well-populated RGB, but F6's RGB is more sparely populated.  
Moreover, the mean color of bright RGB stars in F2 and F9 appear slightly bluer than that of F6.
This means that in its faint outskirts M101 is still asymmetric, at least at $R = 45 \sim 50$ kpc.

\subsection{Radial number density profiles of resolved stars}

Figure \ref{fig_rdp4} shows radial number density profiles of detected point sources in the three CMD bins with distinct ages.
We corrected profiles using the photometric completeness data for each star we derived in Section 2.2.
We divided the profiles into three groups according to azimuthal angles: one on the east side (circles), another one on the west side (squares), and the third on the northeast side (triangles) of M101.

The upper panels of Figure \ref{fig_rdp4} show the number density profiles for the young (MS/HeB), intermediate-aged (AGB), and old (RGB) populations, as selected from the boxes shown in Figure \ref{fig_cmd1}, without the additional background subtraction.
Approximated background levels measured from the control fields (see Appendix \ref{sec:appendix-controlfields} for details) are indicated by a shaded region in each panel.
The profile for the young stellar populations (top left) shows no measurable excess on the top of the background beyond $R \approx$ 40 kpc, while that of the AGB population (top middle) shows a marginal excess at $R \approx$ 45 kpc (F2 and F9).
The excess is clear in the RGB profile where all four data points between $R=40$ and 55 kpc are above the background level (top right).

In the lower panels of Figure \ref{fig_rdp4}, we display the number density profile for each population after background subtraction.
We subtracted the background levels for the ACS and WFC fields separately to get more accurate measurements of the M101 stellar densities.
The effective number projected densities range from $\rho \approx 10^3$ $N/arcmin^2$ to below zero. 
These negative values occur due to low number statistics (a few stars per field) giving rise to large uncertainties and are shown as $1\sigma$ upper limits only.
The extended RGB profile gives a significant detection above $\rho = 1/arcmin^2$ out to $R \approx$ 55 kpc and a marginal detection thereafter.
We used the RGB profile as an indicator of surface mass profiles at the outskirts of M101 in the following analysis.

\begin{figure}
\centering
\includegraphics[scale=0.8]{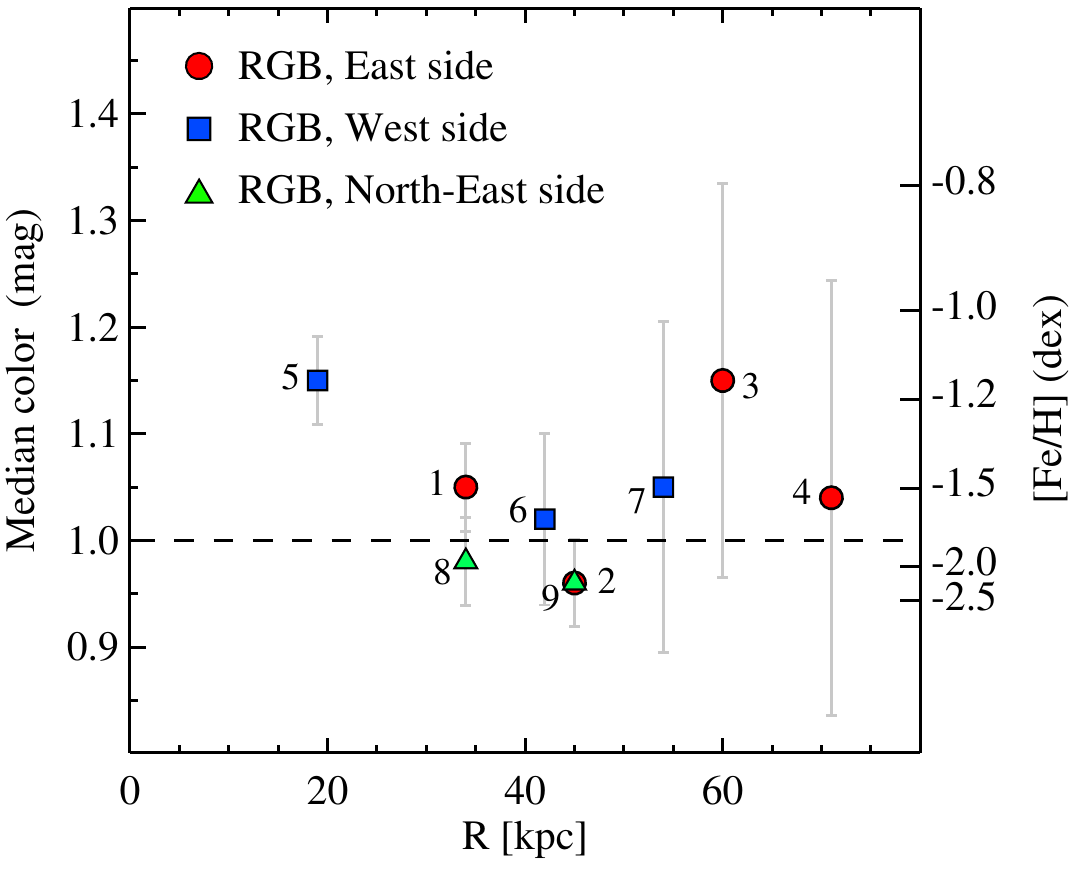}
\caption{Median $(F606W-F814W)_0$ color profiles of RGB stars along the east (circles), west (squares), and northeast (triangles) sides of M101.
The mean RGB color of the fields beyond 40 kpc is indicated by a dashed line. 
}
\label{fig_col}
\end{figure}

\subsection{Radial color profiles of resolved stars}

The colors of the bright old RGB stars are known to have relatively little dependence on age compared to metallicity \citep{hoy55, san66}, so they are often used as a metallicity indicator \citep{har99, mon13, gil14, lee16}.
We investigate the radial color distribution of the RGB stars to get an insight into the metallicity profile of the outer parts of M101.

We selected stars brighter than $F814W_0 = 26.1$ mag, corresponding to $\sim1$ mag below the TRGB. 
The boundary of the selection box is shown by red lines in Figure \ref{fig_cmd1}.
We used both blue and red RGB stars to minimize the selection bias in the color measurements.
The faint stars with $F814W_0 \geq$ 26.1 mag were not used in the analysis due to their relatively low completeness with large color uncertainties.
We transformed the $(F606W - F814W)_0$ color of each selected star to the $Q$-index, a modified color for the RGB stars that is relatively insensitive to the F814W band magnitude \citep{mon13}.
The $Q$-index is defined by rotating CMDs an angle of  $-8\fdg29$, so that a sequence of metal-poor RGB stars becomes vertical. 
The $Q$-index can be expressed by a numerical form, 

\begin{equation}
Q = (F606W-F814W)_0+ F814W_0/6.7 
\end{equation}

where a value of 6.7 is a correction factor determined from a stellar model for old (10 Gyr) and metal poor ([Fe/H] = --1.2 dex) RGB stars.
A detailed description of the $Q$-index can be found in \citet{mon13, mon16}.\\

We determined the median $Q$-index color and corresponding uncertainty using the bootstrap resampling method.
We constructed CMDs of M101 fields subtracted for the contribution of foreground stars and unresolved background galaxies statistically using CMDs of the empty fields displayed in Appendix~\ref{sec:appendix-controlfields}.
The remaining stars were used for an initial guess of the median $Q$-index color.
We iterated this procedure one thousand times to get a distribution of the measured $Q$-index colors. 
Fitting the Gaussian model to the distribution of the colors, we obtained the median color (the Gaussian mean) and corresponding $1\sigma$ uncertainty (the Gaussian width).
We also included a systematic uncertainty due to the photometric calibration of 0.04 mag in color \citep{rad11, mon16}.
Once we measured the median $Q$-index color, we rotated back to the original coordinate and derived a median $(F606W - F814W)_0$ color at 0.5 mag below the TRGB ($F814W_{0,-3.5}$ = 25.6 mag). 
By the definition of the $Q$-index, the median $(F606W - F814W)_0$ color of M101 is 3.82 mag (25.6/6.7) smaller than the median $Q$-index color.
Figure \ref{fig_col} shows a distribution of the measured $(F606W - F814W)_0$ colors as a function of galactocentric distance.
We converted the median $(F606W - F814W)_0$ color to metallicity using the empirical relation in \citet{str14} assuming [$\alpha$/Fe] = 0.3 and marked the values in the right side of the figure.

The measured median colors have a narrow range from $(F606W-F814W)_0 \approx$ 0.95 (F2 and F9) to 1.15 (F5 and F3).
The uncertainty becomes larger due to low stellar densities toward the outskirts.
As expected from the CMDs seen in Figure \ref{fig_cmd1}, the color of F5 in the innermost region is clearly redder than most of the other fields.
We noted, however, that the CMD of F5, as well as F8, shows highly populated upper helium-burning stars and some of them are placed in the blue-side of the RGB selection bin in CMDs.
Thus, we argue that the true median colors of of RGB stars in F5 and F8 would be slightly redder than the measured colors.
We determined the median color of M101 outskirts using six fields beyond 40 kpc (F2, F3, F4, F6, F7, and F9), where the contribution of the young stellar population is not significant.
The weighted mean of the median color of the six fields gives $(F606W - F814W)_0 = 0.98\pm0.03$ with a standard deviation of 0.09 mag.
Here the color measurement is not based on some fields that have only a small number of stars, but on a combination of both sparse and well populated fields taking the weighted mean.
We adopted a conservative color, $(F606W - F814W)_0 = 1.00\pm0.05$.
At this color, the mean metallicity is estimated to be [Fe/H] $=-1.8^{+0.3}_{-0.6}$.

\subsection{Radial surface brightness profiles from integrated light studies}

The radial density profiles of the resolved stars can be combined with those of the integrated light, so we can investigate more detailed structural properties of M101.
There are two independent studies of M101 based on the integrated light photometry by \citet{mih13} and \citet{mer16}. 
\citet{mih13} presented a comprehensive analysis of the extended disk using deep $B$ and $M$ ($\approx V$) images taken with the Burrell Schmidt telescope. 
The $1\sigma$ limiting surface brightness of the $B$ band data was $\mu_{B} = 29.5$ mag arcsec$^{-2}$, where a $9 \times 9$ binned image was used (resulting in a pixel scale of $13\farcs1$).
\citet{mer16} carried out a similar study using $g$ and $r$-band data taken with the Dragonfly array.
They also measured the $1\sigma$ limiting surface brightness in $12\arcsec \times 12\arcsec$ boxes and obtained $\mu_{g} = 28.6-29.2$ mag arcsec$^{-2}$, similar to that of \citet{mih13}.

Both studies mentioned above presented radial surface brightness profiles of M101 out to $\sim50$ kpc.
\citet{mer16} presented the azimuthally averaged profile only, whereas \citet{mih13} provided profiles along several azimuthal bins showing a variation of surface brightness at a given radius.
\citet{mih13} pointed out that the M101 disk is strongly asymmetric, showing a variation in surface brightness of $\sim2$ mag arcsec$^{-2}$ outside $R = 5\arcmin$ (9.7 kpc).
Because our \hst\  data cover only a small fraction of the M101 outskirts, the azimuthally decomposed profiles shown in \citet{mih13} would be more suitable for a comparison.

It is also noted, however, that the Dragonfly array is specifically designed to have less scattered light at large radii, typically by a factor of 5 -- 10, than most other telescopes can achieve \citep{abr14,san14}.
This is especially important at the outskirts of M101, where the surface brightness is as low as $\mu \sim 30$ mag arcsec$^{-2}$.
We, therefore, take a hybrid approach, which utilizes the image data provided by the Dragonfly survey and follows the method of \citet{mih13}.

We obtained public $g$ and $r$-band images of M101 from the Dragonfly survey webpage$\footnote{ http://www.astro.yale.edu/dragonfly}$.
The images are coadded and resampled to have a pixel scale of $2\farcs0$.
We performed surface photometry on star-subtracted images with circular apertures.
We first tried to reproduce the profile shown in \citet{mer16} to make sure that our independent analysis reproduces statistically consistent result of the original author.
We found that the star-subtracted images still contain compact sources around M101, which are mostly background galaxies or known satellite galaxies of M101. We masked these sources to sample faint stellar emission at the outskirts of M101 as much as possible.
The sky background was determined from the outer area with galactocentric distances between $25\arcmin$ (48~kpc) to $35\arcmin$ (68~kpc).
We divided this area into 20 subregions as a function of distance and derived the mean sky value of each region. 
The mean and standard deviation of the 20 sky values were used as the mean and uncertainty of the sky background.
We also carried out  point spread function (PSF) fitting photometry using  DAOPHOT \citep{ste87} on star un-subtracted images and derived instrumental magnitudes, which we used in turn to obtain photometric zero-points by comparing to the PSF photometry provided by the Sloan Digital Sky Survey \citep{yor00}.

\begin{figure}
\centering
\includegraphics[scale=0.9]{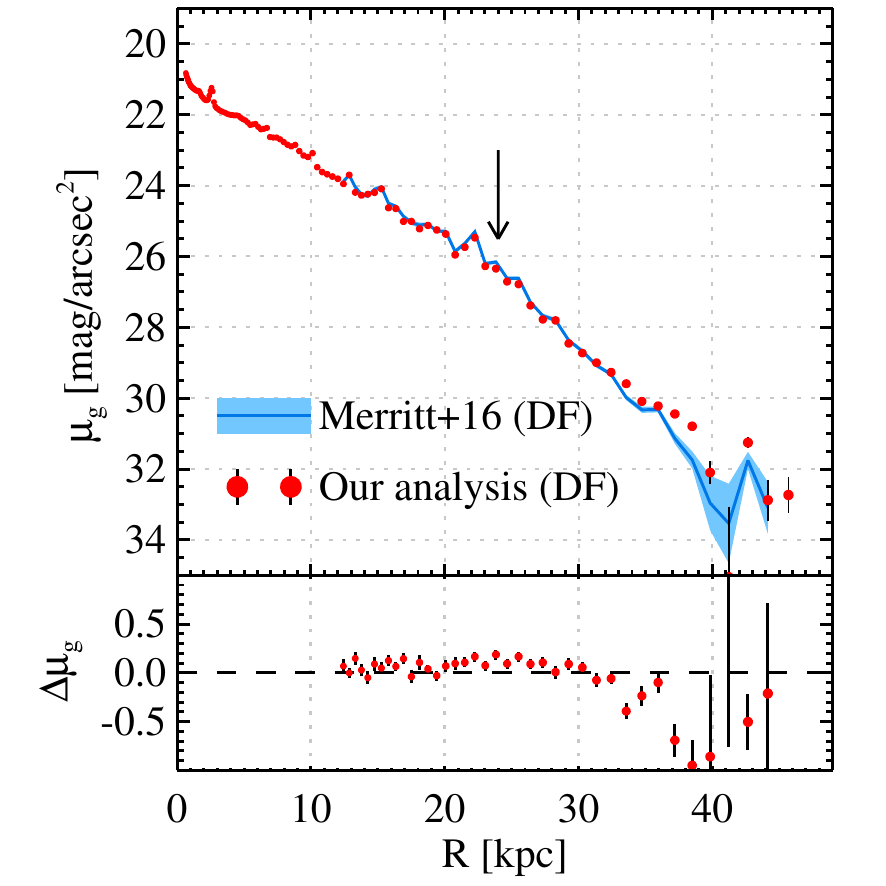}
\caption{(Top) Azimuthally averaged $g$-band radial surface brightness profiles of M101 in \citet{mer16} (solid line with a shaded region) and in this study (red dots with error bars), reproduced from the image data provided by the Dragonfly survey. 
Two profiles agree well.
A weak break in the brightness profile is detected at $R\sim24$ kpc (arrow).
(Bottom) The difference between the two profiles, in the sense that our profile is $\sim 0.08$\,mag fainter than that of \protect\citet{mer16}. 
}
\label{fig_df}
\end{figure}

\begin{figure*}
\centering
\includegraphics[scale=1.6]{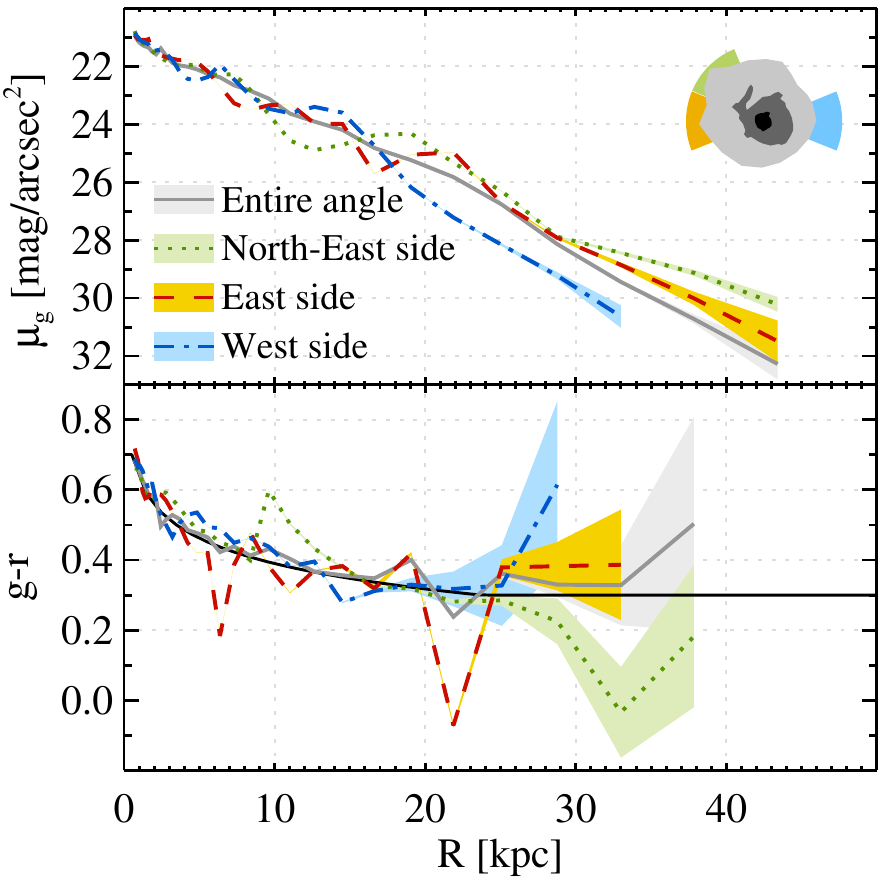}
\caption{(Top) Radial $g$-band surface brightness profiles of M101 taken along different azimuthal selections: azimuthally averaged (solid gray line), along the northeast side (green dotted line), along the east side (orange dashed line), and along the west side (blue dot-dashed line). 
Shaded regions represent uncertainties.
A schematic picture of M101 has the same color coding as the profiles.
Three profiles along the northeast, east, and west sides show large differences at the outskirts.
(Bottom) Radial color profiles. The thick solid line is an approximation of $g-r$ color from the equation 10 of the text. 
}
\label{fig_azi}
\end{figure*}

The top panel of Fig.\ \ref{fig_df} shows an azimuthally averaged $g$-band surface brightness profile of M101 from our analysis (red dots with error bars) and in \citet{mer16} (solid line with a shaded region).
It is clear that the two profiles are in good agreement within uncertainties.
We display the difference between the two profiles in Figure \ref{fig_df}(b), subtracting the profile in \citet{mer16} from ours.
Two points are noted.
First, there is $\Delta\mu_{g}=0.08$ mag arcsec$^{-2}$ systematic offset in the inner region (R $\lesssim$ 30 kpc), where the signal to noise ratio of the residual profile is high. 
The origin of this offset is unclear but may be related to the photometric zero-point adopted in each study.

Second, there is a large offset between the two profiles in the outer region ($R \gtrsim$ 35 kpc), where the surface brightness is fainter than $\mu_g \sim 30$ mag/arcsec$^{-2}$.
The profile in this region is very sensitive to the adopted sky background level, and an unambiguous determination of sky level and its uncertainty are challenging.
We therefore emphasize that the uncertainty we derived from the image and shown in Figure \ref{fig_df} by error bars should be considered as a lower limit.

The azimuthally averaged profile of M101 in Figure \ref{fig_df}, as well as that in Figure 3 of \citet{mer16}, shows a weak break at $R \sim$ 24 kpc ($12\arcmin, \sim5 R_d$).
Beyond this region, the profile becomes steeper, similar to those of the Type II disks \citep{fre70, van79, poh06}.
Our finding is in contrast to that of \citet{mih13}. 
They pointed out that there is no obvious break in the azimuthally averaged disk profiles, though there is a  very slight down-bending at $R = 7\arcmin-9\arcmin$ (14 -- 17 kpc).
This is due to a difference in methodology:
\citet{mih13} derived the averaged profile by taking the median of profiles along several azimuthal bins, while we take the profile of the azimuthally-averaged surface brightness.

We next consider the surface brightness profiles along the northeast, east, and west sides from the center of M101. 
 We selected, from the Dragonfly data, wedge-shaped regions with position angles of $\phi = 45\degree \pm 22.5\degree$ (northeast side, counterclockwise from the north), $\phi = 90\degree \pm 22.5\degree$ (east side) and  $\phi = 270\degree \pm 22.5\degree$ (west side), and derived surface brightness profiles in a similar way as we did for the azimuthally averaged profile.
The background levels were determined locally from the outer area with $25\arcmin \lesssim R \lesssim 35\arcmin$ associated with the target direction.
In Figure \ref{fig_azi} we display $g$-band surface brightness and $g-r$ color profiles of M101 taken along the northeast (green dotted line), east (orange dashed line), west (blue dot-dashed line) sides.
We also display the profile averaged over all azimuths by a gray solid line.

A few distinguishable features are noted.
First, the surface brightness profiles for the northeast, east and west sides are $1.5\sim2$ mag shallower compared to the azimuthally averaged one. 
The brighter limiting magnitudes of the 
azimuthally decomposed profiles
are mainly due to the limited photon flux associated with the small area used for the photometry and for the sky estimation.
Second, there is a large difference between the three azimuthally decomposed profiles, as already mentioned in \citet{mih13}.
The northeast and east sides show radially more extended and brighter profiles outside 18 kpc compared to that of the west side. 
The azimuthally-averaged profile is similar to that of the east side.
Third, the decomposed profiles show larger fluctuations in both surface brightness and color than the averaged profile. 
This is also related to the aperture size used for the averaged photometry. 
The decomposed profiles are more affected by structure (spiral arms, etc.) in the galaxy than the averaged profile. 
Indeed, we confirmed that the sharp blueward dips at $R \sim$ 7 and 22 kpc in the color profile on the east side are due to star forming knots in the spiral arms.
The fluctuation is more evident in color, so we approximate the color profiles using a form below: %

\begin{equation}
(g-r) =
\begin{cases}
    -0.24 \log(R)+0.63       & \quad \text{if } R \leq 24 \text{ kpc}\\
    +0.30,  & \quad \text{if } R > 24 \text{ kpc}.
\end{cases}
\end{equation}

\noindent The approximation is shown in Figure \ref{fig_azi}(b) by a black line and was used to derive stellar mass density profiles in the next section.

\begin{figure}
\centering
 \includegraphics[scale=0.85]{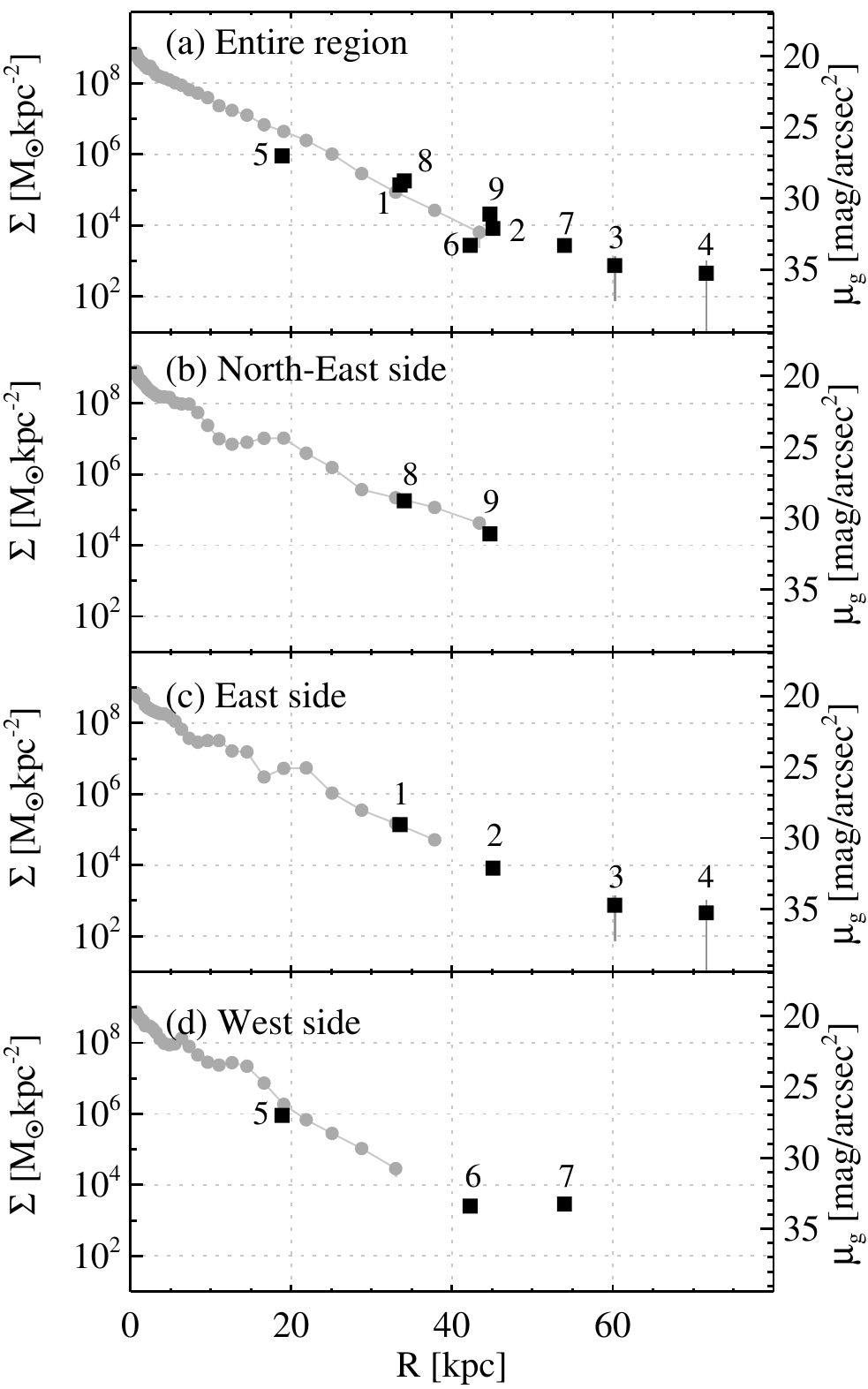}
\caption{Radial mass surface density profiles taken from the entire region (a), the northeast side (b), the east side (c), and the west side (d) of M101.
Circles and squares represent the profiles from the integrated light and the resolved blue RGB stars, respectively. 
Error bars are smaller than the symbol sizes in most cases so they are not readily seen.
}
\label{fig_rdp2}
\end{figure}

\subsection{Radial mass density profiles}

We converted the stellar number density (star count) and the surface brightness (integrated light) profiles to stellar mass density profiles.
We then combined them to see the mass distribution of M101 from inner to outer regions. 
The conversion from stellar number density to mass density was derived using the Padova stellar models \citep{bre12}, similar to the method described in \citet{har17}.
Briefly, we constructed a well-populated CMD for old (10 Gyr) and metal-poor ([Fe/H] $=-1.8$) stars, assuming a Chabrier initial mass function \citep{cha03}.
We then derived a relation between the number of RGB stars 
in the RGB bins of the CMDs (see Fig. 3) and the present-day stellar mass of the entire number of generated stars. 
Here we considered the RGB population only, excluding younger stellar populations, because CMDs of the observed stars 
showed that the outer region of M101 is dominated by metal-poor RGB stars. 
We assign a systematic uncertainty of 0.3 dex to these density estimates, accounting for a reasonable range of age (5 Gyr to 12 Gyr) and metallicity ([Fe/H] = --2.0 to --1.5) variations in stellar populations. 

The conversion between the $g$-band surface brightness and the mass density is made using equation (1) in \citet{van14}:

\begin{equation}
\log(\Sigma) = -0.4(\mu_g-(m-M)_0) + 1.49(g-r)+4.58
\end{equation}

\noindent where $\Sigma$ is stellar mass surface density in $M_{\odot}$ kpc$^{-2}$, and $\mu_g$, $(m-M)_0$, and $g-r$ are measured surface brightness, distance modulus, and color, respectively.
This relation is based on observed galaxies at low redshift ($0.045 < z < 0.055$) with similar stellar masses ($10 < \log(M/M_{\odot}) < 10.7$) and colors ($0.2 < (g-r) < 1.2$) in the SDSS DR7, assuming a Chabrier initial mass function \citep{cha03}.

Figure \ref{fig_rdp2} displays the mass density profiles of M101 taken over 
the entire azimuthal range (a),
the northeast side (b), 
the east side (c), and 
the west side (d).
Circles and squares indicate the profiles from the integrated light and the resolved RGB stars, respectively.
The density profiles extend out to $R \sim 70$ kpc, where the mass densities are as low as $10^3 \mathrm{M_{\odot}~kpc^{-2}}$.
The profiles from the two independent approaches overlap in the region between $R \approx$ 25 kpc to 45 kpc 
and they are in good agreement with each other, especially considering the decomposed profiles only. This agreement between the results for integrated light and star counts is not by design as two completely separate calibrations are used; a color-stellar $M/L$ conversion for the integrated stellar light, and scaling of the RGB star counts to total stellar mass using isochrones. 

There are non-negligible offsets in panel (a) 
such that the mass densities of F5 and F6 are slightly lower and F8, F9, and F2 are slightly higher than those of the integrated light.
These offsets are much smaller in the azimuthally decomposed cases in panels (b), (c) and (d), indicating that the offsets are mainly due to local variations.
The remaining offset between F5 and the integrated light in panel (d) can be explained by the sampling issue, as the star count profile samples the blue RGB stars only, while F5 RGBs have a broad range of color, and also the young stars present at this radius contribute to the integrated light.

The outskirts of the profiles are largely determined by scaling resolved RGB star counts. 
The profile for the M101 west side (d) is almost flat in the outskirts, 
clearly not continuing the exponential disk profile from the inner regions.
This extended stellar structure has a surface brightness of $\mu_g\approx 33$ mag arcsec$^{-2}$ and is similar to stellar halos as seen often in nearby disk galaxies \citep{mcc09, bar12, mon13, oka15, mon16, har17}.
A similar flattening in mass density is seen in the outskirts of the M101 east side (c) but the signal is weaker and the uncertainty large than on the west side.
The profile for the northeast side (b) does not extend far enough in radius to tell whether or not it flattens.

\subsection{Modeling the M101 halo}

The detection of stellar halos in face-on galaxies is very difficult because the distinguishing feature of the accreted populations -- higher velocity dispersion and scale height -- is observationally inaccessible if we only have photometric information. 
We are forced into much less certain inferences based on either mass density profiles or stellar populations.
The case of M101 is even worse because of its extended and asymmetric stellar disk.
Nevertheless, our analysis 
implies that M101 has a stellar halo such that 
the radial mass density profiles become flatter at the outer regions ($R\gtrsim50$ kpc), deviating from the exponential distribution and the stellar populations at the outskirts ($R\gtrsim40$ kpc) are dominated by metal-poor RGB stars. 
From these observational measures, we model the halo structure of M101.
We considered two possible approaches:
one using the radial mass density profiles under the assumption that stellar halos have radially more extended profiles than stellar disks, 
and another from the stellar population perspective assuming that stellar halos of the MW mass galaxies are dominated by metal-poor RGB stars.
In this case, we introduced truncated disk models to explain metal-poor stellar populations  of M101 beyond $R$ = 40 kpc.
The halo geometry of M101 is not known and it is very hard to constrain from the current data so we assumed a spherical geometry in the following analysis.

\subsubsection{Modeling the stellar halo based on the mass density profiles}

We consider a two component model, with an exponential disk and a power-law halo, to express the density profiles of M101:

\begin{equation}
\Sigma(R) = \Sigma_{0,d} {\rm exp}\left(\frac{R}{R_d}\right) +  \Sigma_{0,h} \left(\frac{R}{{\rm kpc}}\right)^{\alpha_h}, 
\end{equation}

\noindent where $R_d$ is the disk scale length, $\alpha_h$ is the projected power-law index for the halo, and $\Sigma_{0,d}$ and $\Sigma_{0,h}$ are surface density scale factors for the disk and halo, respectively.
M101 is widely considered to be a bulgeless pure disk galaxy \citep{kor10}; consequently, we do not include a bulge component in our model. 
We model the outer region of the profile with $R \geq 25$ kpc, where the disk is reasonably well described by a single exponential law and less affected by local variations in the light profile.

\begin{figure}
\centering
 \includegraphics[scale=0.85]{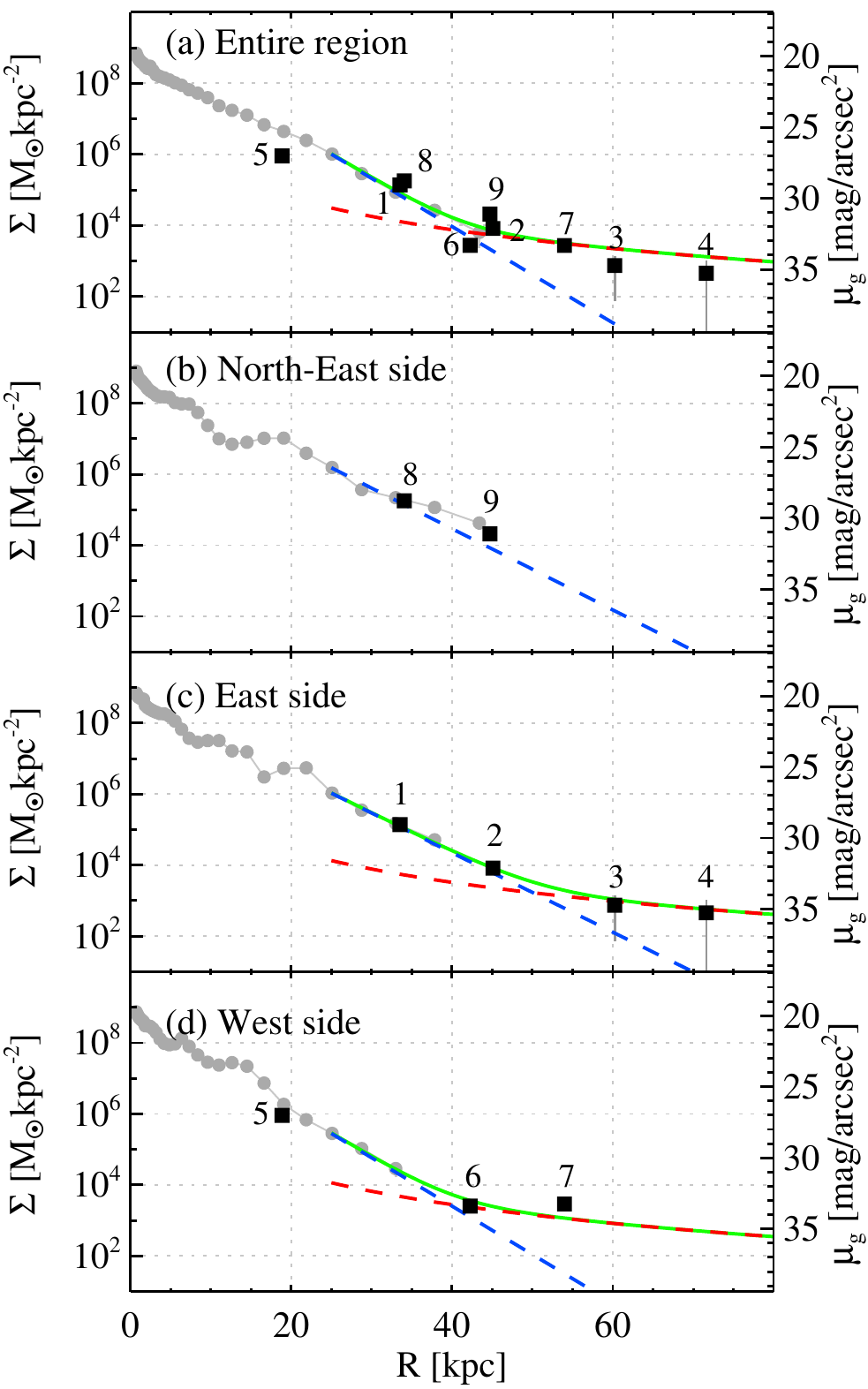}
\caption{
Same as Figure \ref{fig_rdp2}, but with two component fits to the density profiles: outer disk (blue dashed lines, exponential law), halo (red dashed lines, power law). The sum of the two component fits beyond 25 kpc are indicated by green lines. 
The halo component is not detected in the northeast side (b) so we plotted a fit for the outer disk only.}
\label{fig_rdp3}
\end{figure}

We first set all four parameters ($R_d, \alpha_h, \Sigma_{0,d}$, and $\Sigma_{0,h}$) to be free, and fit the outer region of density profiles ($R\geq 25$ kpc) using the IDL based routine {\em mpfitexpr} \citep{mar09}.
The routine yielded the best fit values of the four parameters, but the values were sensitive to initial conditions and errors were very large.
We noted that there are only a few data points beyond $R$ = 40 kpc of the density profiles, so that the parameters for the halo component ($\alpha_h$ and $\Sigma_{0,h}$) are difficult to constrain.
We therefore constrain the power-law index for the halo to a fixed value. While later for error analyses we choose a range of  $\alpha_h=-2$ to $-4$ following the range of $\alpha_h$ values shown by other GHOSTS galaxies \citep{har17}, we for now fix $\alpha_h=-3$ at the mid-point of that range and fit the profiles again. 

The best fit results are shown by dashed lines in Figure \ref{fig_rdp3}. 
The halo component was detected in the profiles for entire region (a), the east side (c) and the west side (d) of M101. 
In the case of the northeast side (b), however, we do not detect the halo component quantitatively. 
This shows that our \hst\  data sampling the outer regions beyond $R \gtrsim50$ kpc is critical in detecting the M101 halo.

\begin{figure}
\centering
 \includegraphics[scale=0.9]{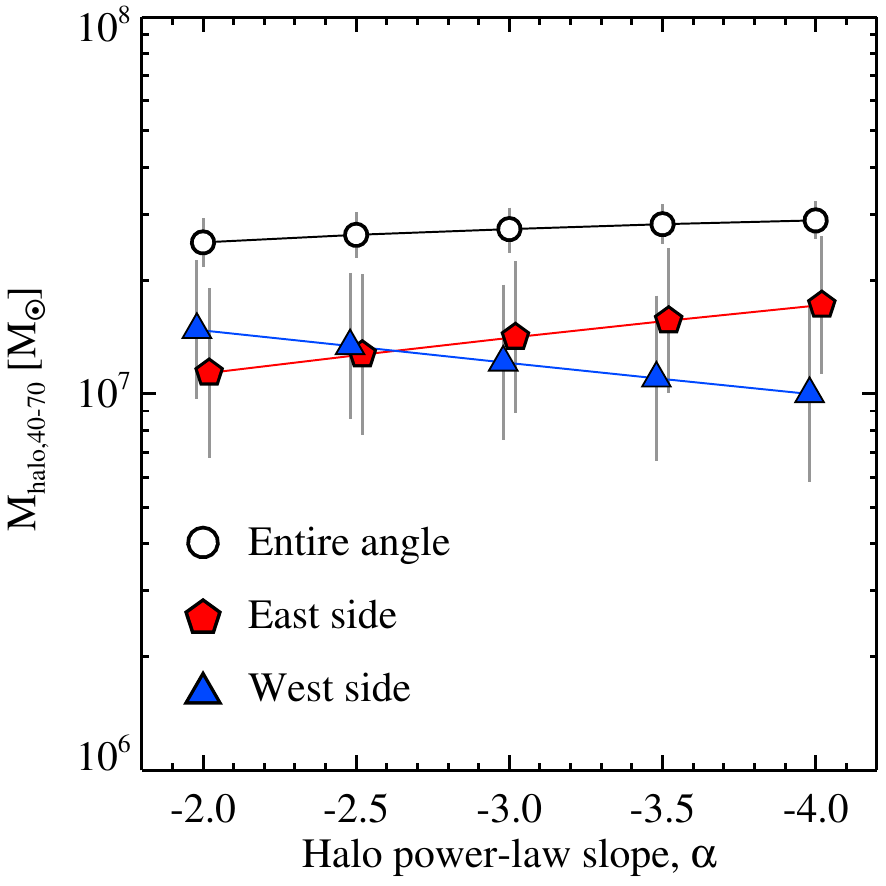}
\caption{
Estimated stellar halo mass between 40 and 70 kpc as a function of the halo power-law slope ($\alpha_h$) assumed in each fit. Circles, pentagons and triangles represent the halo mass from the analysis of the entire region, the east and west sides of M101, respectively.
}
\label{fig_alpha}
\end{figure}

\begin{figure}
\centering
 \includegraphics[scale=0.9]{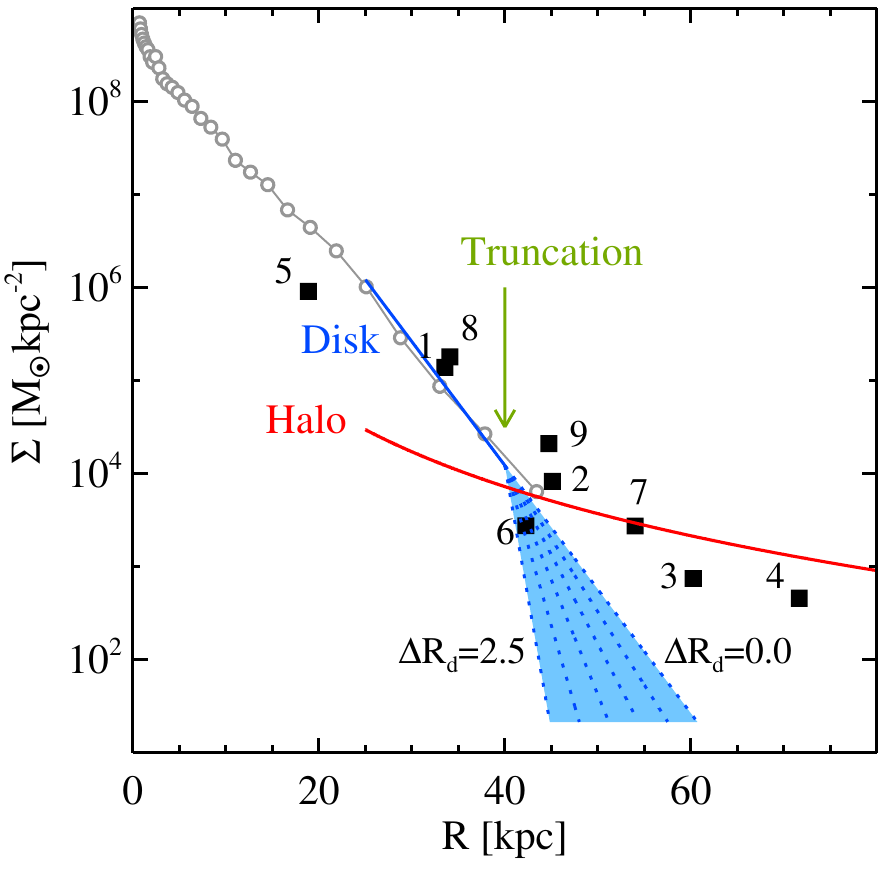}
\caption{Illustration of truncated disk models.
The observed mass density profile for the entire region of M101 is indicated by open circles (integrated light) and filled squares (star counts).
Blue and red lines are models for the disk and halo.
The disk model has a truncation starting at R$\sim40$ kpc with a range of strengths from $\Delta R_d= R_{d,in} - R_{d,out} = 0.0$ kpc (no truncation) to $\Delta R_d=2.5$ kpc (strong truncation).
}
\label{fig_tru}
\end{figure}

\begin{figure*}
\centering
 \includegraphics[scale=0.9]{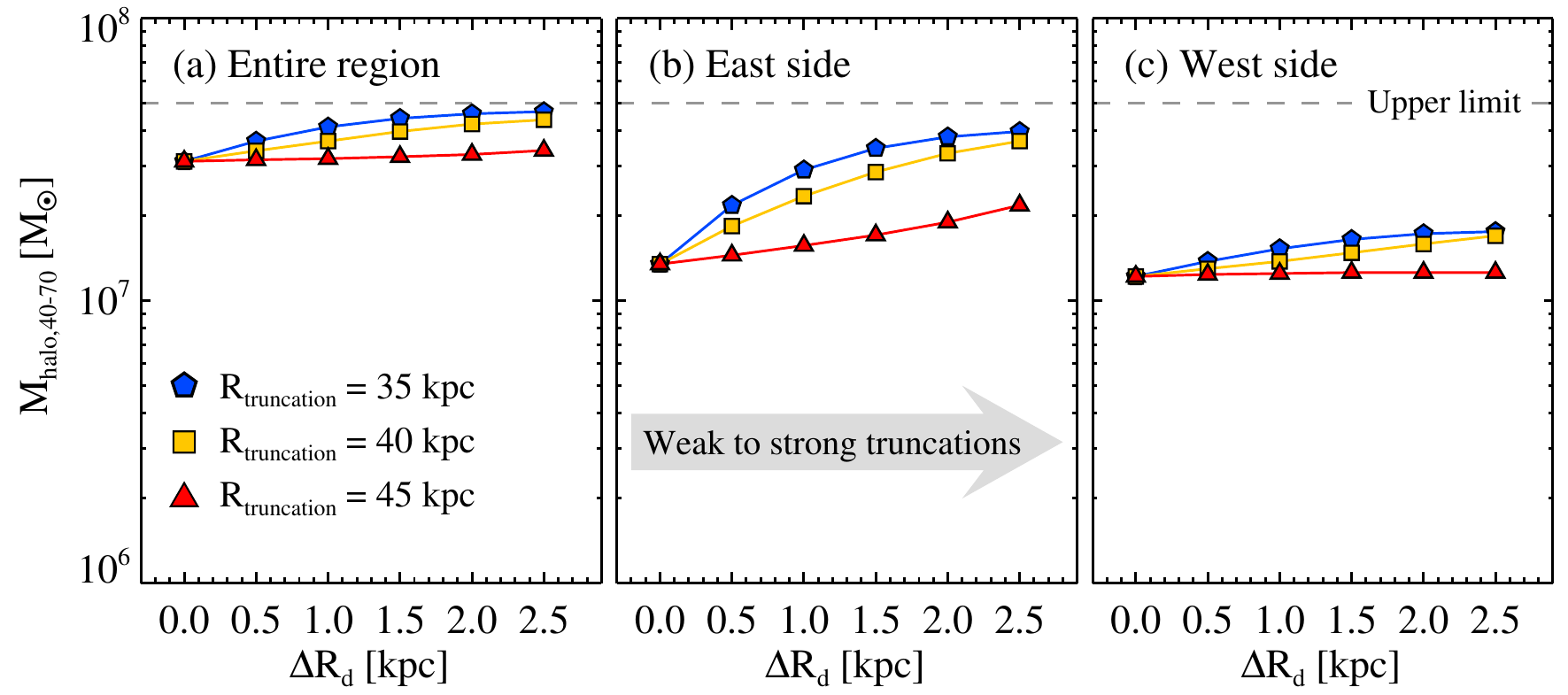}
\caption{
Estimated stellar halo mass between 40 and 70 kpc as a function of the truncation strengths.
Three panels represent results from the analysis of the entire region (a), the east (b) and the west (c) sides of M101.
Disk models with a truncation starting at 35, 40, and 45 kpc are also considered as marked by pentagons, squared, and triangles, respectively.
The upper limit of the estimated halo masses, $M_{halo,40-70} = 5\times10^7 M_{\odot}$, is indicated by a dashed line in each panel.
}
\label{fig_tru_compare}
\end{figure*}

We numerically integrated the halo profile from 40 kpc to 70 kpc, where the contribution of the stellar halo is estimated to be dominant, and derived the stellar halo mass, $\Msub{halo,40-70}$.
The profile for the entire region yields a halo mass of 
$\Msub{halo,40-70} = 2.7^{+0.4}_{-0.4}\times10^7M_{\odot}$.
Here the upper and the lower boundaries are purely  $1\sigma$ fitting errors. 
When we assume that the halo profiles of the east and west sides are good measurements of a spherically symmetric halo, we obtained halo masses of $\Msub{halo,40-70} = 1.4^{+0.8}_{-0.5}\times10^7M_{\odot}$ and $\Msub{halo,40-70} = 1.2^{+0.7}_{-0.5}\times10^7M_{\odot}$, respectively. 
We also confirmed that the measured halo mass on the east side does not change much, when we exclude F4 in the outermost region from the fit ($\Msub{halo,40-70} = 1.5^{+1.0}_{-0.6}\times10^7M_{\odot}$).

%

We next consider halo profiles with different power-law slopes.
\citet{har17} showed that the MW sized disk galaxies have a range of halo power-law slopes from $\alpha_h=$ --2.0 to --3.7.
We set the halo power-law slope $\alpha_h=$ --2.0 to --4.0 in steps of 0.5 and estimated the stellar halo mass, $\Msub{halo,40-70}$, as shown in Figure \ref{fig_alpha}.
The estimated halo masses show only a weak dependence on the power-law slopes assumed.
The mass variations are approximately $\pm15\%$ in a given slope range and are smaller than $1\sigma$ measurement errors.

The estimated halo masses from the entire region of M101 are systematically higher than the values from the east and west sides.
This systematic difference is likely due to the asymmetric disk geometry of M101 combined with the sparse sampling of the galaxy outskirts.
The integrated light profile can be averaged over the entire azimuthal range of the galaxy, while the star count profile is not. 
Modeling the halo using profiles from different spatial sampling could lead to a bias.
This possible bias can be minimized when we use profiles taken along specific directions, where the $HST$ fields are located. 
Indeed, the stellar halo masses measured from the east and west sides show a good agreement in a given slope range.
We determine the halo mass of M101 by taking the weighted mean of the two estimations from the East and West sides with $\alpha_h=-3.0$: $M_{halo,40-70}=1.3_{-0.3}^{+0.6}\times 10^7M_\odot$.
This value is approximately the same in the given range of power-law slopes.
The higher halo mass measured from the entire region ($M_{halo,40-70}\sim3\times 10^7M_\odot$) can be seen as an upper limit, and we consider this in detail in the next section.

The weak mass dependence on the halo profile is encouraging.
The large range in aperture radii used for the mass estimation ($R$ = 40 kpc and 70 kpc) strongly contributes to this relatively small halo mass uncertainty.
The power-law profiles are not very steep in this outer region so that the shape dependent variations in mass can be minimized.
In the following analysis, we used a fixed value of the halo power-law slope, $\alpha_h = -3.0$.

\subsubsection{Modeling the stellar halo from the population perspective}

One of the major advantages of the resolved stellar populations over the integrated light is allowing us to separate the stars into different ages.
CMDs of \hst\  fields in Figure \ref{fig_cmd1} show that there is a transition of stellar populations from inner to outer regions such that inner fields at $R\lesssim40$ kpc (F1, F5, and F8) show multiple stellar populations, 
whereas outer fields at 40 kpc $\lesssim R \lesssim$ 60 kpc (F2, F6, and F7) have no sign of young stars and they are old and metal-poor.
The disk-halo decomposition presented in the previous section suggests that five fields (F1, F2, F5, F8 and F9) are dominated by disk stars, and three fields beyond $R=50$ kpc (F3, F4, and F7), together with a half of F6, belong to the halo component.

Two fields, F2 and F9, are notable. 
They are located at large distances of $R$ = 46 kpc, corresponding to $\sim$10 disk scale lengths, and their CMDs show in general old and metal-poor characteristics (although F9 exhibits a slight excess of AGB stars).
In a stellar population perspective, they could be a stellar halo.
However, they do follow an exponential trend from inner brightness profiles. 
The disk-halo decomposition based on the two simple profiles - a single exponential disk and a power-law halo - suggests that the two fields are solely dominated by disk stars and it would lead to an underestimation of the halo mass.
We modeled the halo structure of M101 with truncated disk models so outer fields ($R\gtrsim$40 kpc) have more weight in the stellar halo.

We considered model disks that have an exponential form with a truncation at $R$ = 40 kpc 
as illustrated in Figure \ref{fig_tru}.
Here, the disk truncation we considered is hypothetical.
There is no observational evidence of a disk truncation at this radius of M101, and it is very difficult to detect, even though it really exists due to the halo component. 
Nevertheless, we explored the possible impact of the disk truncation to estimate an upper limit of the stellar halo mass.
The disk scale lengths before and after the truncation are defined by $R_{d,in}$ and $R_{d,out}$, respectively.
Model disks have different truncation strengths from $\Delta R_d = R_{d,in} - R_{d,out}$ = 0.0 kpc to 2.5 kpc, in steps of 0.5 kpc.
The halo component is assumed to have a power-law slope of $\alpha_h=-3$.
We then fitted the mass density profiles to find  
the disk scale lengths before and after the truncation ($R_{d,in}$ and $R_{d,out}$), and the scale factors for the disk ($\rho_{0,d}$) and halo ($\rho_{0,h}$).
Because the two disk scale lengths ($R_{d,in}$ and $R_{d,out}$) are correlated, there are three unknown parameters to find ($R_{d,in}$, $\rho_{0,d}$ and $\rho_{0,h}$).
The stellar halo mass between 40 kpc and 70 kpc was also calculated by integrating the halo profile numerically as was done for the previous section.
We repeated this calculation for stellar disks with truncations at $R_{tr}$ = 35 kpc and 45 kpc as well. 
Stellar disks with truncations at inner regions with $R <35$ kpc were not considered because CMDs of the inner regions show younger stellar populations.

Figure \ref{fig_tru_compare} shows the estimated stellar halo masses from the truncated disk models as a function of the truncation strength.
We detected the halo component in the profiles for the entire region (a), the east side (b), and the west side (c) of M101.
The halo component was not quantitatively detected in the northeast side even with strongly truncated disk models. 
This is probably due to the lack of data points at a large radial distance ($R \gtrsim 50$ kpc).
It is naturally expected that the inferred halo mass increases when the truncation is stronger and the truncation starts earlier.
A similar trend is seen in the figure.
Disks with strong and earlier truncations yield in general higher halo masses.
We, however, noted that the estimated halo masses are saturated at strongly truncated cases so that they do not exceed the mass of $M_{halo,40-70} = 5\times10^7 M_{\odot}$ (dashed lines).
From this result, we consider a mass of $M_{halo,40-70} = 5\times10^7 M_{\odot}$ as the upper limit of the M101 halo mass.

\begin{figure*}
\centering
 \includegraphics[scale=0.95]{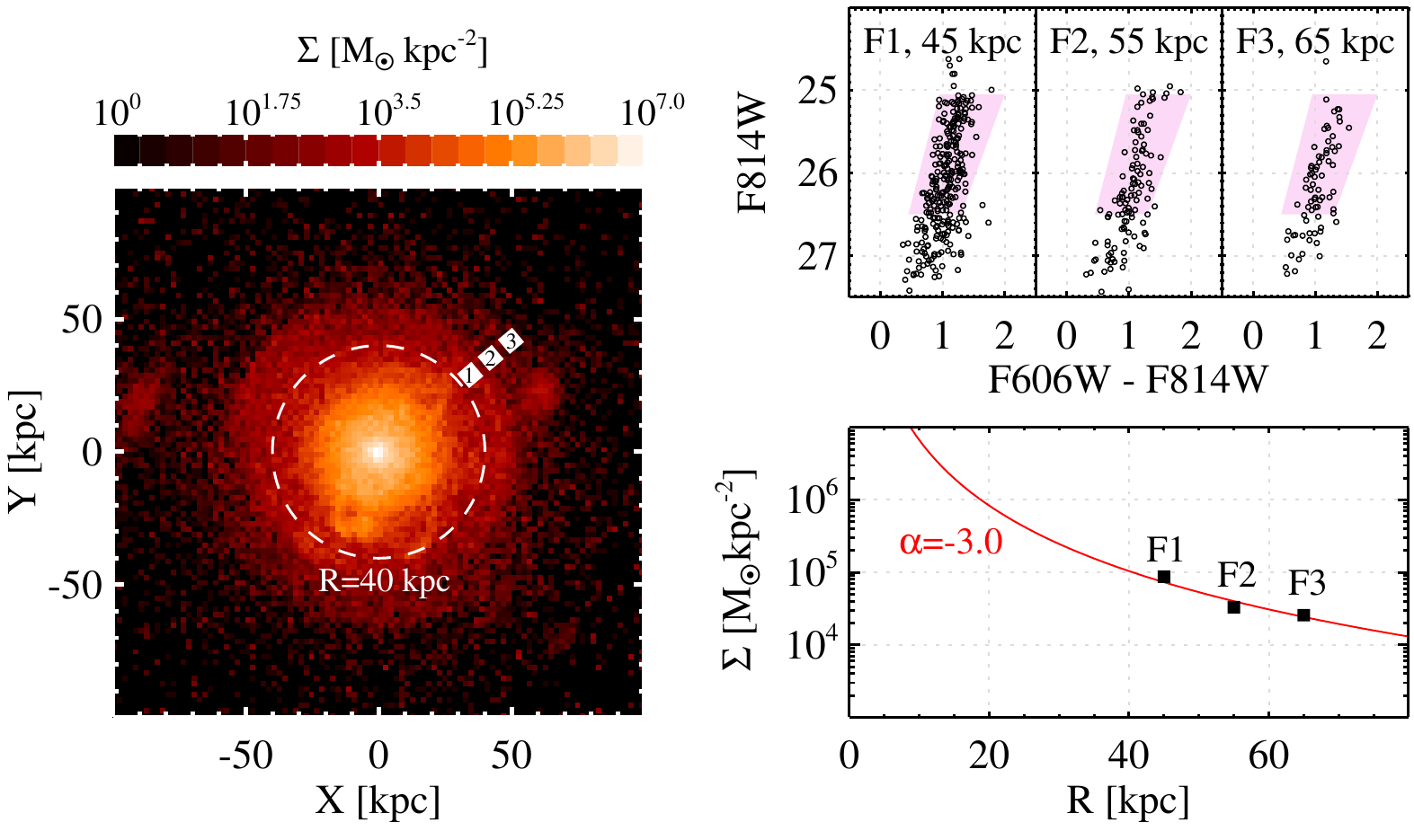}
\caption{
(left) A face-on view of the Halo02 model in \citet{bul05}. 
A dashed line represents $R$ = 40 kpc radius centered on the halo model.
Three ACS-like fields at $R$ = 45, 55, and 65 kpc are marked by white squares.
(top right) CMDs of inferred stars from the three ACS-like fields in the halo model. Stars are shifted to the M101 distance.
Shaded regions are the RGB selection bins used in this study.
(bottom right) Radial mass density profiles measured from the bright RGB stars in the selected fields. 
A solid line indicates the power-law fit with a slope of $\alpha_h=-3.0$.
}
\label{fig_bj05}
\end{figure*}

\subsubsection{Estimating the total stellar halo mass and the halo mass fraction}

Now we have two mass estimates of the M101 stellar halo.
The inference based on the mass density profiles gave the mean stellar halo mass of $M_{halo,40-70}=1.3_{-0.3}^{+0.6}\times 10^7M_\odot$. 
From the stellar population perspective, we examined truncated disk models and obtained an upper limit of the halo mass, $M_{halo,40-70}=5\times 10^7M_\odot$.
These halo mass estimates are based on a limited radial range (40 kpc $\leq R \leq 70$ kpc) with a face-on and circular geometry assumed.
Next we convert the halo masses measured in a limited radial range to total halo masses, which we do by using stellar halo models in a similar way as in our previous studies \citep{mon13, har17}.

\citet{bul05} provide 11 halo realizations from their numerical simulations\footnote{The halo models are available at http://user.astro.columbia.edu/$\sim$kvj/halos/ (see also \citet{rob05} and \citet{fon06}.)}.
They used a hybrid semianalytic plus $N$-body approach to reproduce stellar halos for MW-type galaxies from a series of accretion and disruption events of satellite galaxies in the context of $\Lambda$CDM.
The halo models have a fixed dark matter mass, $M_{vir} = 1.4\times 10^{12} M_\odot$ at z = 0 with a range of stellar halo luminosities from $L = 0.6 \times 10^9$ to $1.2\times 10^9 L_{\odot}$. 
Each halo model has $1.8 \sim 4.6$ million stellar particles that carry basic information, such as positions ($X, Y$ and $Z$), velocities ($V_X, V_Y$ and $V_Z$), age, metallicity, and stellar mass.
We generated a sample of stars having the same age, metallicity, and mass as each stellar particle using the Padova stellar models \citep{bre12}.
We assumed a Chabrier initial mass function \citep{cha03} for the stars in stellar particles.

Figure \ref{fig_bj05} (left) shows an example of one halo realization, Halo02 in \citet{bul05}. A two-dimensional mass density map is plotted in the $XY$ plane so the halo appears face-on, similar to the case of M101.
We considered only the diffuse component of the halo models by selecting the stellar particles that are unbounded from the satellite galaxies ($tub < 0$). 
The mass density map implies that a significant fraction of the mass of accreted stellar halos is located near the center of the host galaxy.

We examined the bias and uncertainty in the derived total stellar halo mass estimates when they are inferred only from the sparse sampling of the outer global structure.
We placed three emulated \hst\  observations at $R$ = 45, 55, and 65 kpc on the model halos as marked by white squares in Figure \ref{fig_bj05}(left).
Each observation has 6 kpc $\times$ 6 kpc size, similar to the mean physical size of the ACS and WFC3 fields at the distance of M101. 
We generated a sample of stars that satisfies age, metallicity, and stellar mass of individual stellar particles in the halo realizations using the Padova stellar models.
CMDs of the generated stars in the three \hst\  observations are presented in the top-right panels of Figure \ref{fig_bj05}.
We shifted stars to the M101 distance and considered observational effects including photometric uncertainties and incompleteness measured from the ASTs, so that the output mock-CMDs appear similar to those from real \hst\  observations. Stars in the mock-CMDs were treated identically to the real M101 stars. We selected stars in the RGB bins of the CMDs and derived mass density profiles as shown in the bottom-right panels of the figure.
The best fit result of a power-law with a slope of $\alpha_h=-3.0$  is plotted by a solid line.
The stellar halo mass, $M_{halo,40-70}$, was derived from the numerical integration of the power-law fit.
We repeated this mass estimate for different halo orientation angles relative to the $X$-axis from $10\degree$ to $360\degree$ in steps of $10\degree$ and for different halo models.
Since we have 11 stellar halo models, we obtained 396 ($= 36\times11$) estimates of $M_{halo,40-70}$.

We found that our analysis technique, employing the two measurements along the opposite sides of the galaxy, induces little bias in mass estimates.
The mean stellar halo mass $M_{halo,40-70}$ inferred from the simulated sparse \hst\  observations is only $2\%$ lower than the true $M_{halo,40-70}$ in the model.
Hence, the expected bias is estimated to be at the level of $\sim2\%$ of $M_{halo,40-70}$, which we ignored in the following analysis.
Our estimates of $M_{halo,40-70}$ are on average  $15.6\%$ of the total stellar mass of the model halos with a standard deviation of $4.7\%$. 
The estimated total stellar halo mass of M101, $M_{halo}$, derived from $M_{halo,40-70}$ and the mean halo mass fraction in simulations ($15.6\%$) is then: 
$M_{halo}=8.2_{-2.2}^{+3.5}\times 10^7M_\odot$. 
We adopted 42\% of $M_{halo}$ as a systematic uncertainty considering uncertainties associated with the uncertain stellar populations (30\%, see Section 3.5) 
and the total halo mass estimation when using sparse $HST$ observations (30\%).
The upper limit of the total halo mass is measured to be 
$M_{halo}=3.2\times 10^8M_\odot$.

The total stellar mass of M101 was derived from its $K$-band luminosity.
The apparent $K-$band magnitude of M101 is 5.51 mag \citep{jar03}, corresponding to an absolute magnitude of $M_K=-23.6$ mag.
Assuming a $K-$band mass to light ratio of M/L = 0.7 
\citep{bel01}, we obtained the total stellar mass of M101, $\Msub{gal} = 4.1\pm1.2\times 10^{10} M_{\odot}$, where the error is dominated by the uncertain stellar populations.
This mass is in agreement with 
\citet{van14} ($5.3\pm1.5\times10^{10}M_{\odot}$) and \citet{mer16} ($5.89\pm1.87\times10^{10}M_{\odot}$).
The stellar halo mass fraction is estimated to be 
$\Msub{halo}/\Msub{gal} = 0.20_{-0.08}^{+0.10}\%$ with an upper limit of 0.78\%.
The measured halo mass fraction is very low but meaningfully higher than zero.

\section{Discussion}
 
Through the analysis of both integrated light and resolved stars, we found a significant evidence for an extended stellar component at the outskirts of M101.
This extended component also shows a clear population-II, old and low-metallicity characteristics indicating that M101 has a stellar halo.
We measured the stellar halo mass at the outskirts of M101 and derived the total stellar halo mass by comparing with the accreted stellar halo models in \citet{bul05}.
In this section, we compare the measured halo properties of M101 with those of previous studies (Section 4.1), those of other MW mass disk galaxies (Section 4.2), and model predictions (Section 4.3).

\begin{figure*}
\centering
 \includegraphics[scale=1.0]{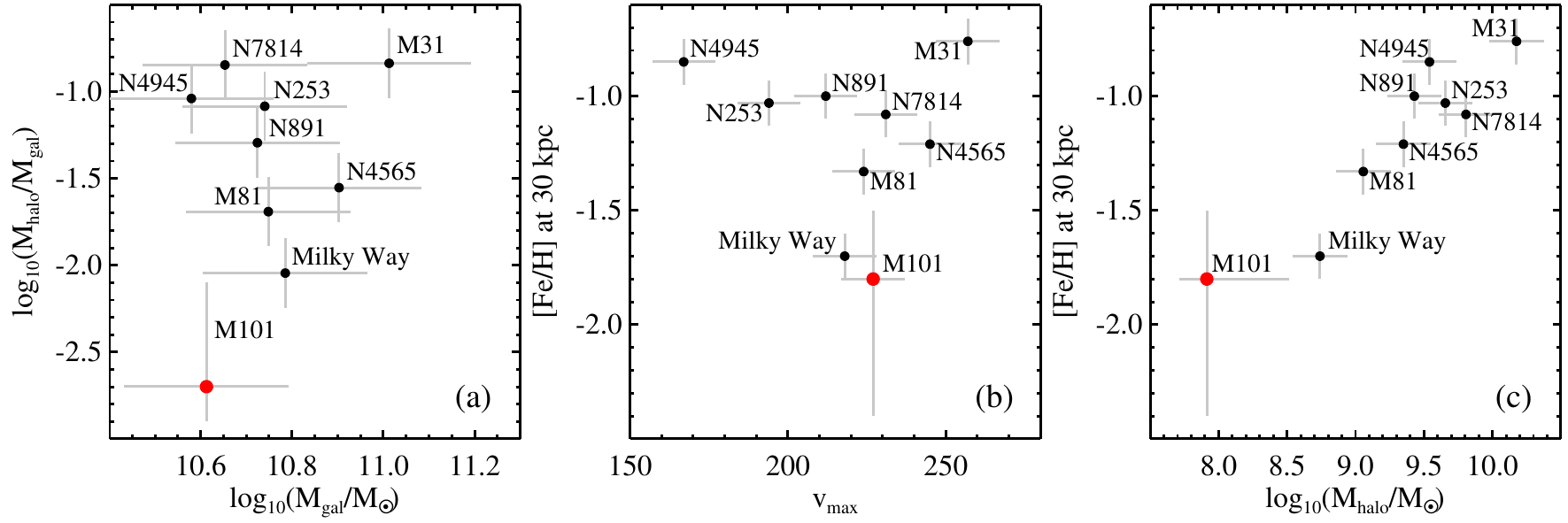}
\caption{Comparison of observed stellar halo properties in nearby galaxies.
(a) Ratio of stellar halo mass and total galaxy stellar mass vs. total galaxy stellar mass.
(b) Mean stellar halo metallicity at 30 kpc vs. maximum rotation velocity. 
(c) Mean stellar halo metallicity at 30 kpc vs. stellar halo mass. 
}
\label{fig_comp}
\end{figure*}

\subsection{Comparison with previous studies of the M101 halo}

There have been several studies investigating the faint outskirts of M101.
The claim that M101 might lack a stellar halo was first made by \citet{van14}.
They used deep $g$ and $i$-band images taken from the Dragonfly array and derived a radial mass density profile reaching out to $R\sim 70$ kpc (see Figures 2 and 3 of their paper). 
The profile appears very steep in the inner 5 kpc region and to follow an exponential profile out to $R \simeq$ 45 kpc. 
Beyond this radius, the profile becomes flatter, deviating from the exponential trend.
This extended structure could be the stellar halo, but its low surface brightness ($\mu_g \gtrsim 32$ mag arcsec$^{-2}$) 
and corresponding large uncertainties ($\sigma \mu_g \gtrsim 1$ mag arcsec$^{-2}$) 
makes it difficult to quantify the halo structure.
They fitted the mass density profile with a two-component model consisting of a \citet{ser68} bulge and an exponential disk.
The residuals from the fit were considered to be a halo component. They fitted the residuals again with a model "U" in \citet{cou11}, which has a power-law shape.
The total stellar halo mass and the stellar halo mass fraction they derived were $M_{halo}=1.7^{+3.4}_{-1.7}\times 10^8M_\odot$ and $M_{halo}/M_{gal} = 0.3^{+0.6}_{-0.3}\%$, respectively.

The stellar halo of M101 has been revisited by the Dragonfly team in \citet{mer16}.
They reanalyzed the M101 data with a revised method to be consistent with eight other galaxies studied in the paper.
The radial surface brightness and mass density profiles of M101 appear to follow an exponential trend out to the end of the profile ($R \sim 50$ kpc). 
They carried out the bulge, disk, and halo decomposition in a similar way as in \citet{van14}.
The stellar halo mass was computed by integrating the halo profile outside 5 half-light radii (32.7 kpc), yielding a halo mass fraction of $M_{halo,>5R_h}/M_{gal} = 0.04\pm0.08\%$. We corrected this value to the total halo mass fraction using the halo models in \citet{bul05}. 
We confirmed that the 11 halo models in \citet{bul05} have $39\pm11\%$ of the total stellar halo mass outside $R = 32.7$ kpc. This mass fraction leads to the total stellar halo mass fraction of M101 of $M_{halo}/M_{gal} = 0.1\pm0.2\%$.

The total stellar halo mass fractions in the two previous studies are very low and not much different from zero within uncertainties.
In this study, we used additional resolved star data as well as the integrated light images to get a better estimate of the halo properties and obtained the stellar halo mass fraction, 
$\Msub{halo}/\Msub{gal} = 0.20_{-0.08}^{+0.10}\%$. 
This value is similar to 
the values in the two previous studies based on the integrated light only.
However, its uncertainty is smaller so that the halo mass fraction is more significant. 
We also provide an upper limit of the halo mass fraction, 0.78\%, from the stellar population perspective.
If this value is taken, the halo mass fraction of M101 is not much different from the that of the Milky Way ($\sim1\%$, \citet{har17}).

\citet{mih18} used two very deep \hst\  fields at $R=35$ and 46 kpc from the M101 center to study the stellar populations of M101's outer disk. 
These two fields are the same as Fields F8 and F9 used in this study. 
They noted that the stellar contents in the two fields, which are only $\sim10$ kpc away from each other along the radial direction, are remarkably different.
The inner field exhibits multiple stellar populations, while the outer field shows no sign of a young stellar population. 
The mean metallicity of the RGB stars in the inner and outer fields they measured are $[M/H] = -1.3\pm0.2$ and $-1.7\pm0.2$, respectively. 
The apparent metallicity difference between the two fields may imply that the stars in the fields have different origin.
Based on the old and low-metallicity characteristics of the outer field, they inferred that the outer field samples the halo stars. 
However, they also pointed out that the disk-halo discrimination based on the stellar populations alone is not easy, so they claimed that it was still unclear whether M101 has a stellar halo or not.

The outer field in \citet{mih18} (F9 in this study)
lies close to the extended stellar disk of M101 (NE plume). 
It makes it difficult to rule out the possibility that F9 is contaminated by disk stars, 
especially due to the presence of bright AGB stars.
The mass density of F9 shown in Figure \ref{fig_rdp3} in this study is not much different from the exponential fit, supporting the disk origin.
It is worth comparing it to fields F2 and F6, which also lie at approximately the same galactocentric distance ($R \sim 43-46$ kpc), but sample different azimuthal directions from the M101 center. 
The CMDs of these fields are dominated by old and metal poor RGB stars with a weaker sign of the bright AGB stars.

The origin of stars in F9 is still unclear because of its compatible properties such that its stellar population is close to the stellar halos, but its stellar density is more similar to the stellar disks.
Given the dominant population of old and metal-poor RGB stars and the presence of very bright AGB stars in F9, it is likely that F9 is mostly halo, but that there might be some low-level star formation that happened in the last Gyr, or some small contribution from disk stars.
If we assume that F9 is dominated by the disk stars, based on its stellar density only,
then the outer disk stellar populations become very similar to the halo.
If instead we consider the stellar populations as a better probe so that F9 becomes the halo, then M101's stellar disk is truncated somewhere between F8 (35 kpc) and F9 (46 kpc).
Further observations beyond F9 would be useful to decompose the stellar disk and halo components in the density profile and to investigate the asymmetry in the disk and halo structures of M101.

\subsection{Comparison of the M101 stellar halo with other observed stellar halos}\label{sec:comparison-otherhalos}

Detailed halo properties of six MW mass disk galaxies have been studied through the GHOSTS survey \citep{dej07, mon16, har17}. 
Now we have one more sample galaxy, M101, which 
has a stellar halo with a very low mass fraction and low metallicity,
making the stellar halo properties of the MW mass galaxies even more diverse.
In this section, we compare the stellar halo properties of M101 to those of the six other GHOSTS disk galaxies, as well as the MW and M31.

We obtained the stellar halo properties of the six GHOSTS galaxies from \citet{har17}.
The properties of the MW and M31 are available in the literature, and well summarized in \citet{har17}.
Briefly, we adopted the total stellar mass of the MW and M31 of 
$\Msub{MW} =  6.1\pm1.1 \times 10^{10} M_{\odot}$ \citep{lic15} and 
$\Msub{M31} = 10.3\pm2.3 \times 10^{10} M_{\odot}$ \citep{sic15}, respectively.
Rotation velocities are estimated to be 
$V_C=218\pm6$ km s$^{-1}$ for the Milky Way \citep{bov12} and 
$V_C=257\pm6$ km s$^{-1}$ for M31 \citep[HyperLeda,][]{mak14}.
The halo metallicity at 30 kpc for the MW is [Fe/H] = --1.7 \citep{ses11, xue15}. 
This value is much lower than that of M31, [Fe/H] = --0.5 \citep{dso18}. 
The stellar halo mass for the MW and M31 are estimated to be $\Msub{halo,MW}  = 4-7 \times 10^8M_{\odot}$ \citep{bla16} and  $\Msub{halo,M31} = 1.5 \times 10^{10}M_{\odot}$ \citep{iba14}, respectively. 
Further details can be found in section 7.1 of \citet{har17}.

In panel (a) of Figure \ref{fig_comp}, we compare 
the ratio of stellar halo mass to the total galaxy stellar mass as a function of the total galaxy stellar mass.
The total galaxy stellar masses range from $\log_{10}(\Msub{gal}/M_{\odot}) = 10.6$ to 11.0, varying by a factor of $\sim2.5$.
The stellar halos show much larger variations in mass fractions by a factor of
$\sim7$ for the previous six GHOSTS galaxies alone, 
$\sim16$ for the six GHOSTS galaxies and two galaxies in the Local group (the MW and M31), 
and more than $50$ for all nine galaxies including M101.
M101 shows extreme properties in the total galaxy mass and the stellar halo mass fractions among the sample galaxies.

In panel (b) of Figure \ref{fig_comp}, we plot 
the stellar halo metallicity at 30 kpc as a function of maximum rotation velocity, which is a proxy of the total galaxy mass.
The halo metallicity of M101 is measured at $R\sim50$ kpc, because inner $R = 30$ kpc regions are not accessible due to the disk stars.
The radial color profile of RGB stars in Figure \ref{fig_col} does not show a clear gradient beyond $R = 40$ kpc.
We estimated the halo metallicty at 30 kpc of M101 from the metallicity at 50 kpc ($[Fe/H]=-1.8^{+0.3}_{-0.6}$) assuming that there is no metallicity gradient between 30 kpc and 50 kpc.
Instead, we assigned an additional metallicity uncertainty of 0.2 dex for M101 considering the mean color gradient of the GHOSTS stellar halos, $\Delta Q = -0.002$ mag/kpc \citep{mon16}.
It is worth mentioning that the halo metallicities for the six GHOSTS galaxies were measured along the minor axis, the vertical axis away from the disk, while the value for M101 is from the major axis, which is only accessible.
In the figure, we see that there is no measurable trend.
The stellar halo of M101 is similar to the MW in rotation velocity and metallicity. 

Panel (c) of Figure \ref{fig_comp} displays the 
stellar halo metallicity at 30 kpc as a function of inferred stellar halo mass. 
The two halo properties show a strong correlation 
as reported and also discussed in detail in \citet{har17}. 
The stellar halo of M101 follows this correlation, extending the low mass end.
This correlation is broadly consistent with expectations in cosmological simulations that most of the halo mass is accreted from the few most-massive progenitors \citep{vog14, dea16, bel17, dso18}.
The low stellar halo mass and halo metallicity of M101 together with the MW indicate that these two galaxies had amongst the quietest accretion histories in the Local Universe.

\begin{figure*}
\centering
\includegraphics[scale=0.45]{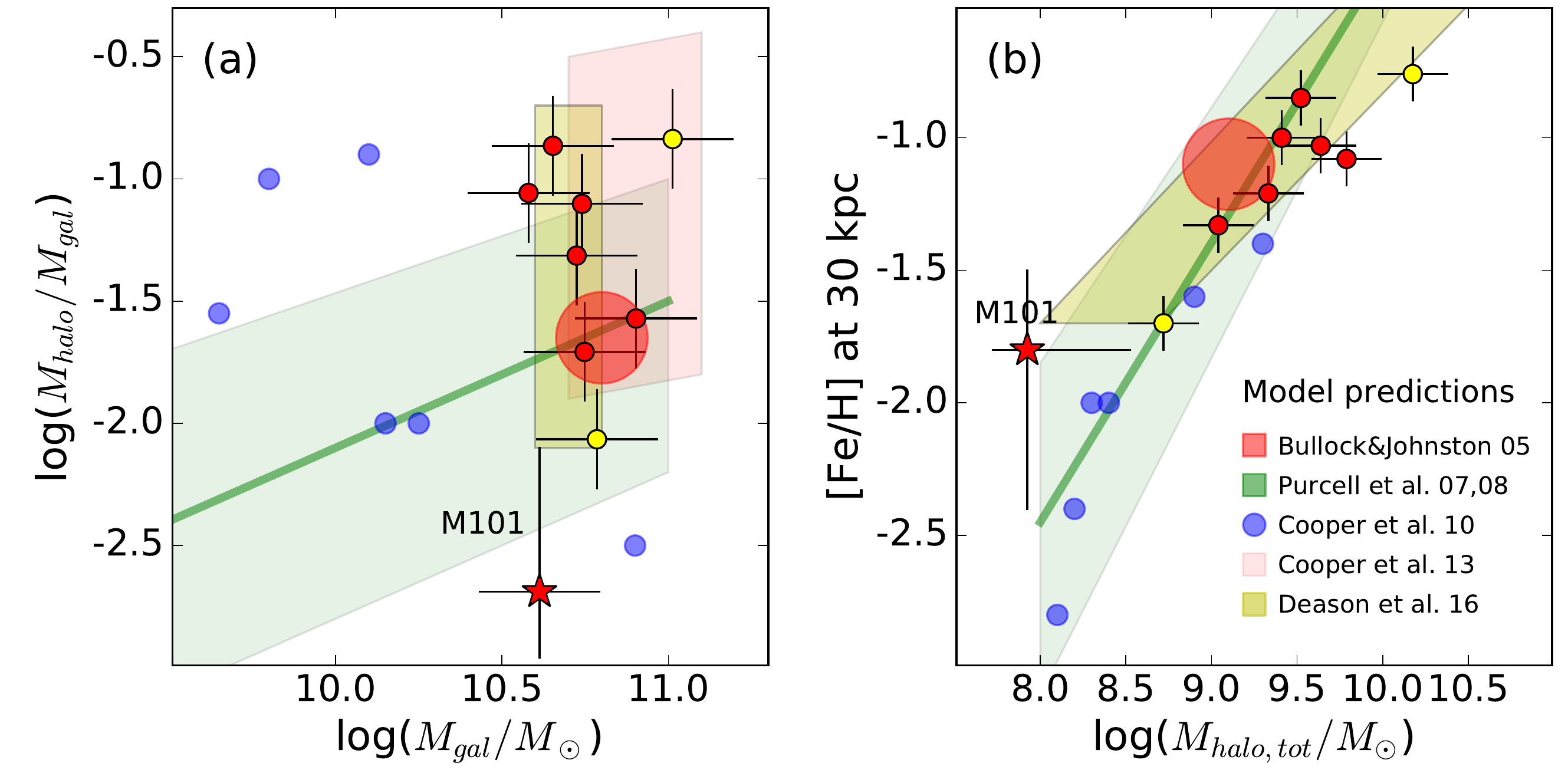}  
\caption{(a) Total stellar halo mass fraction as a function of total galaxy stellar mass.
(b) Stellar halo metallicity as a function of the total stellar mass of the galaxy.
Observed galaxies are marked by symbols with error bars : 
six GHOSTS galaxies (red circles), 
the MW and M31 (yellow circles), and 
M101 in this study (starlet symbol).
Shaded regions indicate model predictions in the literature : \citet[][red area]{bul05}, \citet[][light green area with a green line]{pur07, pur08}, \citet[][blue circles]{coo10}, \citet[][pink area]{coo13}, \citet[][light yellow area]{dea16}. 
}
\label{fig_model}
\end{figure*}

\subsection{Comparison of the M101 stellar halo with models}

In Figure \ref{fig_model}, we explore the stellar halo properties of observed galaxies in the context of model predictions accounting for the 
accretion of low-mass dwarf galaxies during halo formation.
We derived model predictions of stellar halos compiled in \citet{har17} and plotted their approximated boundaries in the figure by shaded regions: 
\citet[][red area]{bul05}, 
\citet[][light green area]{pur07, pur08}, 
\citet[][blue small circles]{coo10}, 
\citet[][pink area]{coo13}, 
\citet[][light yellow area]{dea16}.
Observed galaxies are also marked by filled symbols with error bars.

As already mentioned in Section~\ref{sec:comparison-otherhalos}, the observed stellar halos have a broad range of the total stellar halo mass fractions ($\Msub{halo}/\Msub{gal}$). 
This diversity is also found in the model predictions in panel (a), as well as in \citet{mon19}, where this is also found with the Auriga hydrodynamical simulations.
M101 has the lowest stellar halo mass fraction among the observed galaxies. The measured halo mass fraction of M101, together with its upper limit, marks the low-mass end of the \citet{pur07,pur08} models (within 95\% boundary) and also agrees with the general trend of the \citet{coo10} models.
A strong correlation between the stellar halo mass ($\Msub{halo}$) and metallicity ([Fe/H]) is well reproduced by models, as shown in panel (b).
The M101 stellar halo marks the low mass and the low metallicity end of the observed halos, but nicely agrees with the model predictions.

The overall agreement of the M101 halo with models is quite encouraging.
Clearly M101 has a low mass and low metallicity stellar halo compared with those of other disk galaxies; such a halo occurs naturally in current halo formation models as the outcome of a relatively quiet accretion history dominated by the accretion of few to several low-mass, low-metallicity satellite galaxies \citep{dea16,dso18, mon19}.  

\section{Summary}

Using deep image data taken with HST/ACS and HST/WFC3, we examine the photometric and structural properties of M101 outskirts.
We obtain resolved star photometry from these high-resolution images and confirm that the outskirts of M101 are dominated by old and metal poor RGB stars.
The radial mass density profiles of M101 show evidence for an extended structural component in M101's outskirts, a strong candidate for being the stellar halo.
We compared the basic properties of this extended structural component with those of other disk galaxies.
Our primary results are summarized as follows.

\begin{enumerate}

\item We derived deep CMDs of resolved stars, reaching down to $\sim2$ magnitude below the TRGB. 
The CMDs beyond 40 kpc show clear Population II, old and low metallicity characteristics.

\item The radial star count profile of the Population II RGB stars reaches out to $\sim60$ kpc. The mean metallicity of the RGB stars at R $\sim50$ kpc is estimated to be [Fe/H] $=-1.8^{+0.3}_{-0.6}$.

\item We independently derived radial surface brightness profiles of M101 using public image data provided by the Dragonfly team and combined them with our resolved stellar photometry to derive radial mass density profiles.

\item The radial mass density profiles of the M101 east and west sides show an extended stellar component, which we interpret as being the stellar halo.
In the case of the northeast side, however, we do not detect the halo component quantitatively.

\item We derived the halo mass of M101 from the exponential disk + power-law halo model fits and obtained a halo mass of $M_{halo}=8.2_{-2.2}^{+3.5}\times 10^7M_\odot$, and a halo mass fraction of $\Msub{halo}/\Msub{gal} = 0.20_{-0.08}^{+0.10}\%$.
In this case, the outer disk of M101 is assumed to have very old and metal-poor stellar populations, similar to halo populations.

\item We also estimated an upper limit of the halo mass fraction, $\Msub{halo}/\Msub{gal} = 0.78\%$, from the assumption that 
stellar halos can be traced better by the old and metal-poor populations.
This modeling favors a truncated disk at $R \sim 40$ kpc to explain the population change.

\item We compared the halo properties of M101 with those of other disk galaxies. From the mass, mass fraction, and mean metallicity of the M101 halo, we conclude that M101 has a low-mass, low-metallicity stellar halo with properties similar to that of the Milky Way.

\item M101's low-mass and low-metallicity stellar halo is in agreement with the range of halos expected for Milky Way mass galaxies, where M101 would have had a rather quiet accretion history dominated by the accretion of a few (but more than two) low-mass, low-metallicity satellite galaxies.

\end{enumerate}

\begin{acknowledgements}
We acknowledge discussions with Richard D'Souza. This work was supported by HST grant GO-13696 provided by NASA through a grant from the Space Telescope Science Institute, which is operated by the Association of Universities for Research in Astronomy, Inc., under NASA contract NAS5-26555. Additionally, some of the data presented in this paper were obtained from the Mikulski Archive for Space Telescopes (MAST). We are grateful to the Dragonfly team for obtaining and releasing the excellent quality deep images of M101 that we used in this study.
We are also thankful to the anonymous referee for his/her useful comments.
This research has made use of the NASA/IPAC Extragalactic Database (NED) which is operated by the Jet Propulsion Laboratory, California Institute of Technology, under contract with the National Aeronautics and Space Administration.
This work was supported by NSF grant AST 1008342 and HST grants GO-11613 and GO-12213 provided by NASA through a grant from the Space Telescope Science Institute, which is operated by the Association of Universities for Research in Astronomy, Inc., under NASA contract NAS5-26555.
A.M. acknowledges partial support from CONICYT FONDECYT regular 1181797.
\end{acknowledgements}

\appendix{}

\section{CMDs of the Control Fields and Background Estimation  }\label{sec:appendix-controlfields}
We describe here the method used to estimate the background levels used to determine our radial number density profiles.
We first selected \hst\  fields in the archive satisfying the following criteria: 
(1) located at high galactic latitude with $|b| \gtrsim 40\degree$, similar to M101's $b \sim 59.8\degree$, 
(2) located at large distance from bright stars or galaxies in the sky, 
(3) located at large distance from known tidal streams or star clusters in the galactic halo,
(4) located at large distance from known galaxy clusters, and
(5) observed with similar exposure times as the M101 data ($\sim 2500s$).
Table \ref{tab_obs2} summarizes the selected \hst\  fields taken with ACS (8 fields) and WFC3 (8 fields) from the selection criteria above. 
Some of the fields have been used in \citet{rad11} and \citet{mon16} to determine the best selection criteria (photometry culls).

\begin{figure*}
\centering
\includegraphics[scale=0.9]{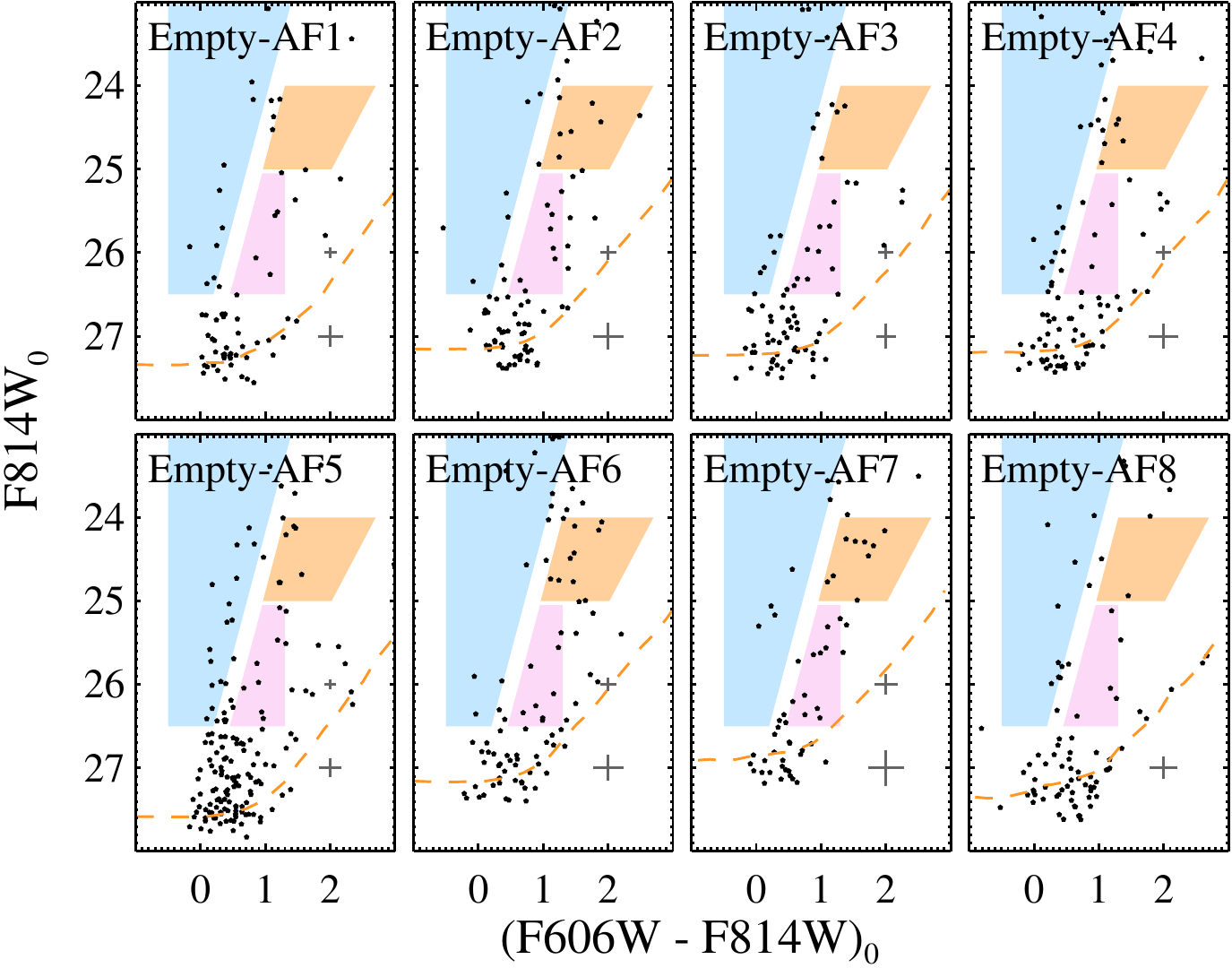}
\caption{CMDs for the selected point sources corrected for the extinction and reddening and after the culls have been applied in the eight empty fields taken with ACS/WFC. Shaded regions are the same as those in Figure \ref{fig_cmd1} : Blue for young stellar population including main sequence and helium burning stars, orange for intermediate aged AGB stars and red for old RGB stars. 
Dashed lines represent the 50\% completeness limits.
Photometric errors as derived from ASTs at $(F606W - F814W)_0$ = 1.0 are marked at $(F606W - F814W)_0$ = 2.0.
}
\label{fig_cmd_acs}
\end{figure*}

\begin{figure*}
\centering
\includegraphics[scale=0.9]{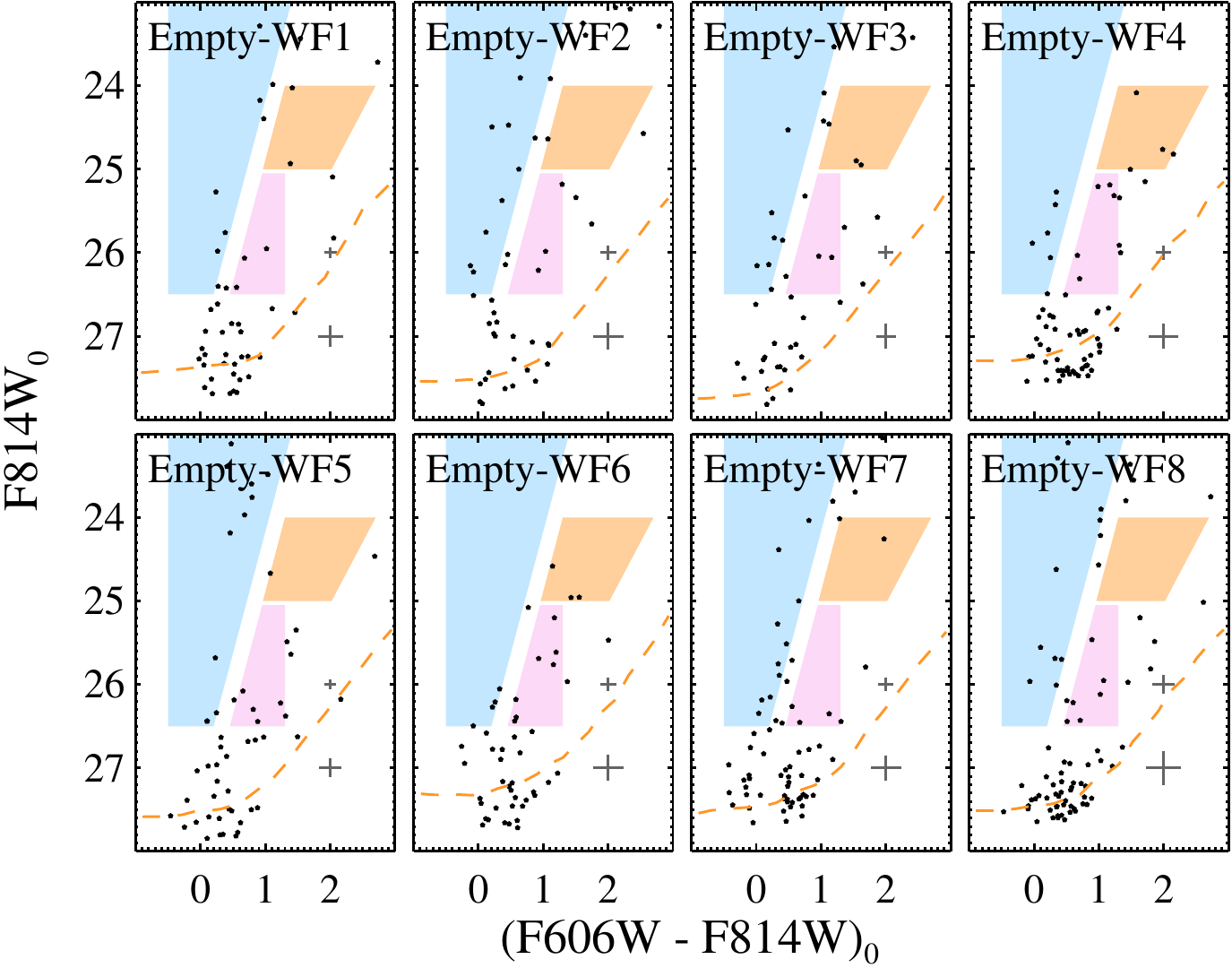}
\caption{Same as Figure \ref{fig_cmd_acs}, but for WFC3 fields.
}
\label{fig_cmd_wfc}
\end{figure*}

We calibrated and extracted point source photometry the \hst\  data for the 16 ``empty'' fields in the same way as the M101 fields, and plotted their CMDs in Figures \ref{fig_cmd_acs} (for ACS fields) and \ref{fig_cmd_wfc} (for WFC3 fields).
Most of the sources in the CMDs must be either foreground stars or unresolved background galaxies and AGN that have passed the culls.
Shaded regions in each figure denote the population bins used for the M101 fields.
Sources are sparsely distributed in the CMDs without showing remarkable structures.
We counted the number of sources in the population bins of each CMD and calculated the spatial number densities considering the photometric incompleteness and the effective field area. Derived results are summarized in Figure A.3. 
We confirmed that all populations show field to field variations in spatial number densities. 
We estimate the mean number density of sources in the ACS fields,
$\rho = 0.95\pm0.31$ $N/\mathrm{arcmin}^2$ in the MS + HeB bin,
$0.44\pm0.31$ $N/\mathrm{arcmin}^2$ in the AGB bin, and 
$0.73\pm0.26$ $N/\mathrm{arcmin}^2$ in the RGB bin of the ACS fields, 
where the uncertainties are taken to be the standard deviation of the individual densities.
We obtained the similar values from the WFC3 fields, 
$\rho = 1.08\pm0.40$ $N/\mathrm{arcmin}^2$ in the MS + HeB bin,
$0.20\pm0.14$ $N/\mathrm{arcmin}^2$ in the AGB bin, and 
$0.60\pm0.26$ $N/\mathrm{arcmin}^2$ in the RGB bin.
The measured source densities from the ACS and WFC3 fields agree well within uncertainties.
We used these values for the background estimation.

\begin{table*}
\caption{Summary of HST data for the control fields}
\centering
\begin{tabular}{lrrrrrrr}
\hline
\hline
Field & R.A. & Decl & $b$ & Inst. & \multicolumn{2}{c}{Exposure Time(s)} & Prop ID  \\
 & (2000.0) & (2000.0) & ($^{\circ}$) & & $F606W$  & $F814W$ \\
\hline

Empty-AF1 & 01 05 33.92 &--27 37 03.0 &--86.8& ACS	& 4,080 & 4,080 & 9498 \\ 
Empty-AF2 & 14 17 02.01 &  52 25 02.7 &  60.1& ACS	& 2,260 & 2,100 & 10134 \\ 
Empty-AF3 & 14 19 18.17 &  52 49 24.6 &  59.6& ACS	& 2,260 & 2,100 & 10134 \\ 
Empty-AF4 & 14 21 35.89 &  53 13 48.3 &  59.0& ACS	& 2,260 & 2,100 & 10134 \\ 
Empty-AF5 & 14 25 15.26 &  35 38 20.5 &  68.3& ACS	& 4,500 & 4,500 & 10195 \\ 
Empty-AF6 & 10 56 31.88 &--03 41 18.2 &  48.5& ACS	& 2,160 & 2,160 & 10196 \\ 
Empty-AF7 & 12 38 18.65 &  62 20 51.0 &  54.7& ACS	& 2,092 & 2,092 &  13779 \\ 
Empty-AF8 & 03 32 43.58 &--27 47 55.9 &--54.4& ACS	& 2,556 & 2,556 & 11563 \\   

\hline 
Empty-WF1 &  4 41 48.87 &--38 05 08.5 & --41.0 & WFC3	& 2,500 & 2,500 & 13352 \\  
Empty-WF2 &  8 53 59.34 &  43 52 07.2 &   40.0 & WFC3	& 4,400 & 4,400 & 13352 \\  
Empty-WF3 &  1 25 36.40 &--00 00 47.5 & --61.7 & WFC3	& 3,200 & 6,400 & 13352 \\  
Empty-WF4 & 15 00 22.92 &  41 28 03.9 &   60.0 & WFC3	& 2,650 & 2,000 & 14178 \\  
Empty-WF5 & 10 51 21.54 &  20 21 27.1 &   61.8 & WFC3	& 4,050 & 3,950 & 14178 \\  
Empty-WF6 & 12 08 35.15 &  45 44 20.5 &   69.5 & WFC3	& 3,230 & 3,180 & 14178 \\  
Empty-WF7&  3 33 23.16 &--40 57 27.8 & --54.1 & WFC3	& 3,150 & 3,150 & 14178 \\  
Empty-WF8&  9 47 02.82 &  51 26 12.5 &   47.8 & WFC3	& 2,750 & 2,400 & 14178  \\ 

\hline
\end{tabular}
\label{tab_obs2}
\end{table*}

\begin{figure*}
\centering
\includegraphics[width=0.4\textwidth]{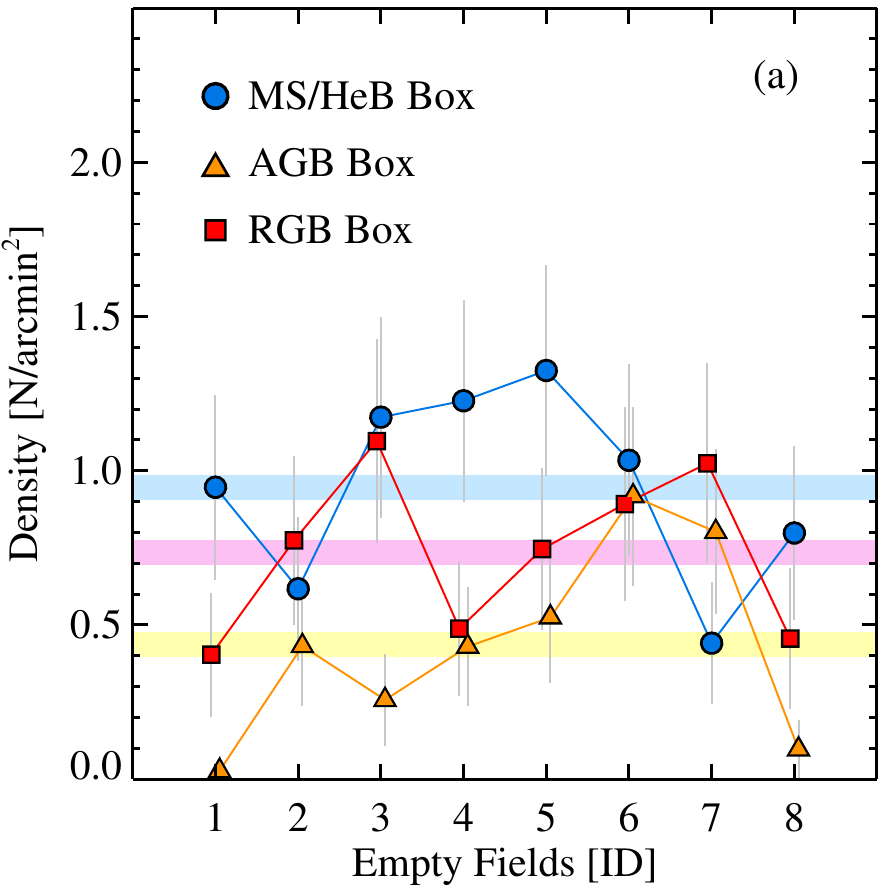}
\includegraphics[width=0.4\textwidth]{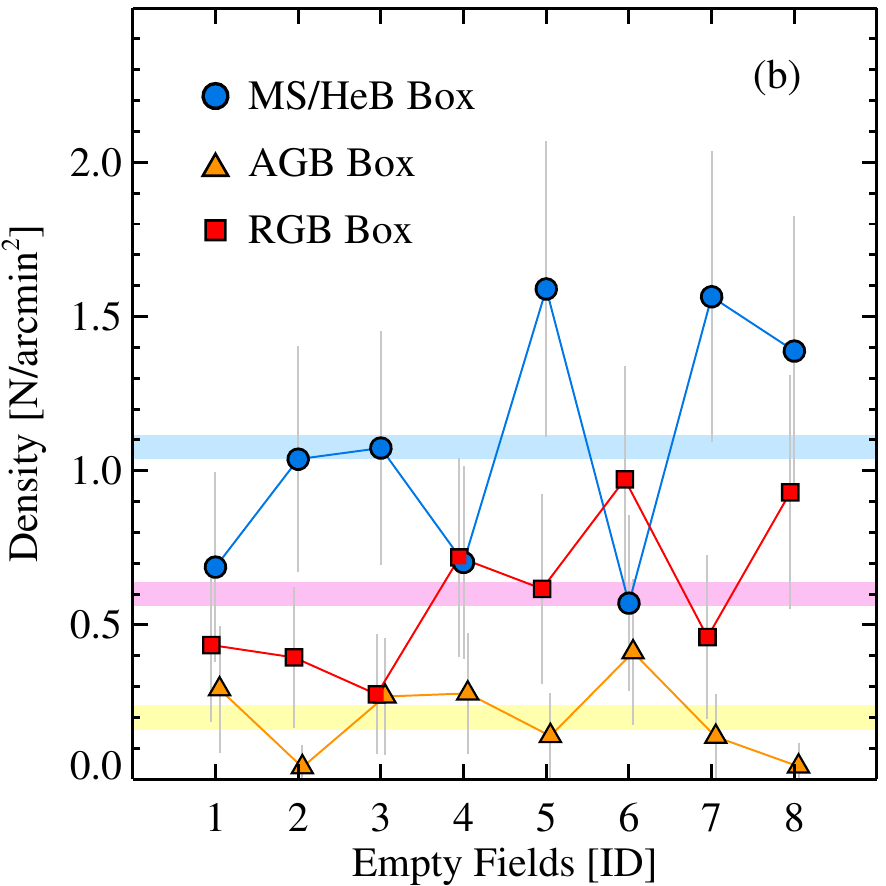}
\caption{(a) Distributions of the unresolved object densities in the MS+HeB (circles), AGB (triangles), RGB (squares) bins in the eight empty fields taken with ACS. Error bars denote $1\sigma$ Poisson uncertainties of the object count.
The mean level of each object population is indicated by a shaded region.
(right) Same as (a) but for WFC3 fields.
}
\label{fig_bg_acs}
\end{figure*}

\end{document}